\documentclass[biblatex]{biophys-new}
\usepackage{xcolor}
\usepackage[colorlinks,allcolors=cyan!70!black]{hyperref}

\usepackage{amssymb}
\usepackage{siunitx}
\title{Quantifying the biophysical properties of stomatocytes in health and disease}
\runningtitle{Quantifying the biophysical properties of stomatocytes} 

\author[1,*]{Zhaojie Chai}
\author[2]{Jianlu Zheng}
\author[3]{He Li}
\author[2,*]{Ming Dao}
\author[1,*]{George Em Karniadakis}
\runningauthor{Chai et al.} 

\affil[1]{Division of Applied Mathematics, Brown University, Providence, Rhode Island, United States}
\affil[2]{Department of Materials Science and Engineering, Massachusetts Institute of Technology, Cambridge, Massachusetts}
\affil[3]{College of Engineering, University of Georgia, Athens, Georgia, United States}

\corrauthor[*]{zhaojie\_chai@brown.edu(ZC);mingdao@mit.edu(MD);george\_karniadakis@brown.edu(GEK)}
\affil[ ]{Zhaojie Chai and Jianlu Zheng contributed equally to this work}

\papertype{Article}

\begin{document}

\begin{frontmatter}

\begin{abstract}
Hereditary stomatocytosis (HS) comprises red blood cell (RBC) disorders characterized by cup-shaped erythrocytes that share a nominal morphology yet respond oppositely to splenectomy: the procedure is often curative in overhydrated HS (OHS) and intermediate phenotypes but can precipitate life-threatening thromboembolism in dehydrated HS (DHS/xerocytosis). This ``splenectomy paradox'' persists because clinicians cannot predict which patients will benefit. A central obstacle is that RBC biomechanics is governed by several partly independent parameters---membrane shear modulus ($\mu$), bending rigidity ($k_c$), surface-to-volume ratio ($S/V$), and cytoplasmic viscosity ($\eta_{\mathrm{cyto}}$)---that existing assays capture only piecemeal. Here we combine dissipative particle dynamics (DPD) simulations with microfluidic imaging to construct a control discocyte (CTR-RBC) and three stomatocyte models (ST-RBC1--3) at fixed membrane area ($A_0 = 132.9~\mu\mathrm{m}^2$) and decreasing volume ($V = 109.7, 101.5, 89.8$~fL), spanning the OHS-to-DHS hydration range. Tracing this single parameter set through five mechanically orthogonal assays, we find that interendothelial-slit (IES) traversal is geometry-dominated: overhydrated ST-RBC1 requires roughly an order of magnitude higher critical pressure than CTR-RBC, whereas dehydrated ST-RBC3 passes freely. ST-RBC3 nonetheless suppresses membrane tank-treading and raises low-shear whole-blood viscosity by ${\sim}29\%$ at physiological haematocrit, comparable in magnitude to the hyperviscosity measured in Gaucher-disease blood. In silico, a funnel--obstacle chip amplifies these differences into a label-free centerline-offset signal predicted to separate all four RBC types (${\sim}4.5\sigma$ between the extreme phenotypes). Together, these results unite single-cell mechanics, splenic filtration, and hemorheology in one framework, offer a mechanical account of the OHS-versus-DHS splenectomy paradox, and point toward a microfluidic route to pre-operative risk stratification in HS.
\end{abstract}

\begin{sigstatement}
Hereditary stomatocytosis produces cup-shaped red blood cells across a spectrum of overhydrated (OHS), mild, and dehydrated (DHS) phenotypes that share a nominal morphology but respond oppositely to splenectomy---curative in OHS, prothrombotic in DHS. Using dissipative particle dynamics coupled to microfluidic imaging, we show that splenic interendothelial slits selectively retain low-surface-to-volume OHS cells, while stiff DHS cells pass freely yet, once in circulation, suppress membrane tank-treading and raise low-shear whole-blood viscosity by $\sim 29\%$. The same underlying mechanics are amplified into a distinctive trajectory signature in a funnel--obstacle microfluidic chip, enabling label-free discrimination of all four RBC subtypes. Our framework resolves the splenectomy paradox in hereditary stomatocytosis and provides a mechanical blueprint for pre-operative risk stratification in RBC rigidity disorders.
\end{sigstatement}

\end{frontmatter}

\section*{Introduction}

Red blood cells (RBCs) are highly specialized, deformable cells whose primary function is to transport oxygen through the microcirculation. To sustain this function throughout their $\sim$120-day lifespan, human RBCs undergo on the order of $10^{5}$ large, reversible deformations as they traverse capillaries and splenic interendothelial slits (IES) that are substantially narrower than their $\sim 8~\mu$m resting diameter~\cite{Mebius2005Structure,Li2018Mechanics}. A healthy discocyte negotiates $\sim 3~\mu$m capillary-scale constrictions and $1$--$2~\mu$m IES without membrane fragmentation, a mechanical feat that underlies stable blood rheology across more than three decades of physiological shear rate and ensures effective tissue perfusion in the microcirculation~\cite{tomaiuolo2014biomechanical,Chien1970Shear,baskurt2003blood}. Loss of this remarkable deformability is a common mechanical denominator across hereditary and acquired RBC disorders, including hereditary spherocytosis and elliptocytosis~\cite{boltonmaggs2011guidelines}, sickle cell disease~\cite{sahun2025novel}, malaria, diabetic microangiopathy, Gaucher disease, and hereditary stomatocytosis~\cite{gallagher2017disorders,andolfo2025evolving,franco2013abnormal,flatt2009hereditary,delaunay1999hereditary}. Because all of these conditions compromise the same composite shell--cytoplasm system but through different molecular routes, a mechanics-first framework for interpreting their flow-dependent consequences has broad translational value.

The mechanical flexibility of RBCs arises from their composite membrane architecture: a fluid lipid bilayer is coupled to a spectrin--actin cytoskeleton, forming a viscoelastic shell (surface area $A_{0} \sim 135~\mu\mathrm{m}^{2}$, shear modulus $\mu_{0} \approx 4$--$9~\mu$N/m, bending rigidity $k_{c,0} \approx 2$--$3\times10^{-19}$~J) enclosing a hemoglobin-rich cytoplasm of effective viscosity $\eta_{\mathrm{cyto}} \approx 5$--$15~\mathrm{mPa\cdot s}$~\cite{dao2021erythrocyte,moreau2023physical,Li2018Mechanics}. Equally important is the geometric reservoir provided by the excess surface area relative to a sphere of the same volume, quantified by the surface-to-volume ratio ($S/V$). For a healthy discocyte ($V \approx 92~\mathrm{fL}$) the $S/V$ ratio is $\sim 1.44~\mu\mathrm{m}^{-1}$, whereas a sphere of equal volume would yield $\sim 1.05~\mu\mathrm{m}^{-1}$; this $\sim 30\%$ reserve membrane area is what permits large shape changes at near-zero area strain. The four parameters $(\mu, k_c, S/V, \eta_{\mathrm{cyto}})$ enter RBC transport in distinct non-dimensional groups: the capillary number $\mathrm{Ca} = \mu_{p}\dot{\gamma}R/\mu$ controls tank-treading and in-plane deformation, the F\"{o}ppl--von K\'{a}rm\'{a}n number $\gamma_{\mathrm{FvK}} = \mu R^{2}/k_{c}$ controls curvature localisation, the confinement ratio $R/h$ determines slit passability, and the sphericity $\Psi$ fixes the minimum channel that the cell can negotiate~\cite{Chien1970Shear,baskurt2003blood,tomaiuolo2014biomechanical,dao2021erythrocyte}. Because each dimensionless group weights the four parameters differently, even modest perturbations of one parameter can translate into disproportionate changes in transit time, clogging probability, or apparent viscosity, which in the microcirculation depends sharply on both hematocrit and tube diameter~\cite{pries1992blood}.

The principal biomechanical checkpoint for RBC integrity is the spleen, which couples two sequential processes: mechanical retention of abnormal RBCs at the IES of the red pulp, followed by phagocytic elimination by resident macrophages~\cite{Mebius2005Structure,Li2018Mechanics,qiang2023microfluidic}. Each day, a healthy spleen surveys the entire circulating pool of $\sim 2.5\times10^{13}$ RBCs, removing only the $\sim 1\%$ that are most senescent or mechanically compromised, consistent with a steady-state RBC lifespan of $\sim 120$~days. \textit{In vivo} RBC transit times at the IES peak near $30$--$60~\mathrm{ms}$~\cite{macdonald1987kinetics}, and direct \textit{ex vivo} measurements on human spleens as well as microfluidic splenon-on-a-chip devices demonstrate that cells with $S/V \lesssim 1.3~\mu\mathrm{m}^{-1}$ or reduced membrane reserve are preferentially retained~\cite{pivkin2016biomechanics,rigatbrugarolas2014splenon}. This retention--elimination balance sets a narrow mechanical window: cells below it accumulate in the splenic cords and shorten their own lifespan, while cells above it populate the systemic circulation and perturb bulk rheology. Diseases that shift this balance asymmetrically therefore produce both hematological (anemia, splenomegaly) and rheological (hyperviscosity, thrombosis) manifestations whose relative severity depends on which side of the balance is disturbed.

Among pathological RBC morphologies, stomatocytes form a distinct class characterized by cup-shaped or coffee-bean-like geometries with a single slit-like stoma, reflecting a deep, asymmetric membrane invagination~\cite{geekiyanage2019coarse,tomaiuolo2014biomechanical}; the classical bilayer-couple hypothesis attributes this sequence to an asymmetry between the inner and outer leaflets of the lipid bilayer, such that a single effective area-difference parameter controls the full stomatocyte--discocyte--echinocyte shape continuum~\cite{lim2002stomatocyte}. Stomatocytes are the hallmark of hereditary stomatocytosis syndromes~\cite{gallagher2017disorders,andolfo2025evolving}, which include overhydrated hereditary stomatocytosis (OHS, $\sim 1/10^{6}$ births, driven by mutations in cation-leak membrane channels), dehydrated hereditary stomatocytosis or xerocytosis (DHS, $\sim 1/50{,}000$, driven by gain-of-function mutations in mechanosensitive cation channels that couple mechanical forces to RBC volume regulation~\cite{cahalan2015piezo1}, together with loss of a Ca$^{2+}$-activated K$^{+}$ channel), and intermediate or mild phenotypes. OHS cells are water-overloaded ($\mathrm{MCV} \sim 110$--$140~\mathrm{fL}$, low $S/V$), nearly spherical, and mechanically fragile; DHS cells are K\textsuperscript{+}- and water-depleted ($\mathrm{MCV} \sim 85$--$95~\mathrm{fL}$, high cytoplasmic haemoglobin concentration $\sim 35$--$40~\mathrm{g/dL}$), coffee-bean-like, and exhibit elevated $\eta_{\mathrm{cyto}}$ and curvature-localized membrane stress~\cite{gallagher2017disorders,andolfo2018genotype}. Despite sharing a stomatocytic morphology, OHS and DHS therefore differ by roughly an order of magnitude in volume perturbation and by nearly an order of magnitude in post-splenectomy thrombosis risk~\cite{picard2019clinical,frederiksen2019dehydrated}, highlighting that ``stomatocyte'' is not a single mechanical phenotype but a family spanning a multidimensional $(\mu, k_{c}, S/V, \eta_{\mathrm{cyto}})$ space.

These mechanical differences translate directly into one of the most striking paradoxes in benign hematology. Splenectomy in OHS and mild stomatocytosis reliably corrects anemia (mean haemoglobin rise $\sim 2$--$3$~g/dL), removing the site of excess retention and prolonging RBC lifespan~\cite{gallagher2017disorders,perrotta2008hereditary}. The same operation is, however, contraindicated in DHS, where post-splenectomy thromboembolism rates have been reported as high as $20$--$40\%$, roughly one order of magnitude above healthy post-splenectomy baselines, despite comparable improvement in haemoglobin~\cite{andolfo2018genotype,andolfo2025evolving,turpaev2025overview,picard2019clinical,frederiksen2019dehydrated,boltonmaggs2011guidelines}. Existing hypotheses invoke elevated cytoplasmic viscosity, loss of endothelial surveillance, and RBC--platelet interactions, but none of these quantitatively connects the patient-specific mechanical phenotype to the observed thrombotic risk. In particular, the relationship between membrane mechanics, single-cell IES traversal, collective clogging, and whole-blood viscosity in stomatocytosis subtypes has not been systematically established. Experimental \textit{in vitro} access to all four observables is limited by the difficulty of isolating individual mechanical parameters, imposing extreme ($\sim 1~\mu$m) confinement, measuring dense population-level dynamics, and following the same cell population from single-cell to suspension scale. Ektacytometry, micropipette aspiration, and microfluidic deformability assays each probe only a subset of the relevant mechanical observables, and clinical hemorheology is confounded by haematocrit, plasma protein composition, and splenectomy status.

Multiscale particle-based modeling has emerged as a complementary route for dissecting mechanics, confinement, and collective flow across molecular, cellular, and tissue scales~\cite{tang2017openrbc,zhang2017multiscale,chai2023dynamics,chai2022periodic,chang2016md,zhang2021deep,zhang2020deep}. Among these approaches, dissipative particle dynamics (DPD) couples a triangulated-network viscoelastic membrane to an explicit mesoscopic plasma, allowing independent control of $\mu$, $k_c$, $S/V$, and $\eta_{\mathrm{cyto}}$ while preserving hydrodynamic interactions at the micrometre scale~\cite{fedosov2010multiscale,Fedosov2011Multiscale,groot1997dissipative,hoogerbrugge1992simulating,toscano2026graftathena}, and it is naturally complemented by finer-grained spectrin-level and optical-tweezer-calibrated RBC models~\cite{dao2003mechanics,li2005spectrin}. DPD has previously reproduced healthy tank-treading~\cite{tran1984determination,williamson1985microrheologic,fischer2007tank} and the rheology of Gaucher-disease blood~\cite{franco2013abnormal,chai2025silico}, and recent two-component DPD studies have begun to map the full stomatocyte--discocyte--echinocyte (SDE) shape landscape and its consequences for capillary viscosity~\cite{wen2025stomatocyte,liu2025capillary,geekiyanage2019coarse}, demonstrating the required quantitative fidelity. What has been missing is a stomatocytosis-focused framework that (i)~spans the full OHS--mild--DHS phenotype range using a single internally consistent parameter set, (ii)~ties simulation to matched single-cell microfluidic imaging, and (iii)~follows the same phenotypes from membrane curvature all the way to whole-blood viscosity and clinical splenectomy risk.

In this work, we build that framework. We use DPD simulations tightly coupled to brightfield microfluidic imaging to construct three stomatocyte models, overhydrated (ST-RBC1), mild (ST-RBC2), and dehydrated (ST-RBC3), that span the experimentally observed morphological spectrum while holding total membrane area fixed at a physiological value ($132.9~\mu\mathrm{m}^{2}$) and cell volume in the range $89.8$--$109.7~\mathrm{fL}$. After anchoring these models to static morphometry and bending-rigidity control of shape, we probe them in one consistent framework through five mechanically orthogonal flow assays: (i)~pressure-driven critical-threshold and dynamic traversal of spleen-mimetic IES, (ii)~single-cell tank-treading under simple shear across $\dot{\gamma}=20$--$200~\mathrm{s}^{-1}$, (iii)~collective clogging in mixed CTR + stomatocyte populations at spleen-like slit arrays, (iv)~shear-dependent whole-blood viscosity over three decades of $\dot{\gamma}$ at physiological haematocrit ($45\%$), and (v)~label-free deformability-based trajectory sorting in a funnel--obstacle microfluidic device (FOR-Chip). By tracing a single parameter set through all five assays, we obtain a unified, quantitative, mechanics-based framework that (a)~resolves the OHS-vs-DHS splenectomy paradox, (b)~predicts the dominant clinical manifestation (anemia vs thrombosis) from the mechanical phenotype, and (c)~translates directly into a microfluidic diagnostic that separates the three stomatocyte subtypes without labels or external forces.

\section*{Materials and Methods}

\subsection*{Dissipative particle dynamics (DPD) model}

We employed the DPD method to simulate the blood plasma as a mesoscopic fluid, following the framework established by Groot and Warren~\cite{groot1997dissipative} and Hoogerbrugge and Koelman~\cite{hoogerbrugge1992simulating}. In this method, the plasma is represented by DPD particles that interact through soft conservative, dissipative, and random forces, which collectively enforce hydrodynamic behavior \cite{Fedosov2011Multiscale}. The dissipative and random forces satisfy the fluctuation--dissipation theorem, ensuring proper thermal equilibration of the system \cite{groot1997dissipative}. The fluid viscosity was calibrated by measuring the shear stress response under steady shear flow, and parameters were chosen to reproduce the physiological viscosity of blood plasma while maintaining numerical stability~\cite{Fedosov2011Quantifying}. In our simulations, the viscosity of the cytoplasm was assumed equal to that of the plasma, which has been shown to adequately capture suspension rheology~\cite{Fedosov2011Predicting}. Thus, of the four mechanical phase-space axes $(\mu, k_c, S/V, \eta_{\mathrm{cyto}})$, the first three are swept across phenotypes (Table~\ref{tab:mechanical_properties}), while $\eta_{\mathrm{cyto}}$ is held at the plasma value for all four cells; variation of $\eta_{\mathrm{cyto}}$ across the physiological MCHC range (corresponding to a $\sim 2$--$3\times$ increase in DHS) is reserved for future work (see Discussion). Similar DPD parameter choices have been shown in prior work to accurately predict blood viscosity over a range of shear rates~\cite{Fedosov2011Predicting}. The specific DPD parameters, including number density, conservative force coefficient, dissipative force constant, and time step, are provided in the Supplementary Material.

The computational domain was periodic in the flow and vorticity directions, with no-slip walls moving at constant velocities to impose shear flow in the suspension simulations. For confined flow simulations, a constant pressure gradient was applied to drive RBCs through capillary-like channels or splenic slit geometries. All simulations were run for sufficient time to reach steady-state after initial transients.

\subsection*{RBC model}

To investigate the role of individual biophysical factors on blood viscosity and stomatocyte-related hemorheological alterations, it is essential to model single RBCs accurately while preserving the underlying hydrodynamics and convective transport processes governing blood flow. A healthy RBC is a highly deformable, nucleus-free biconcave membrane with a characteristic diameter of approximately $8~\mu$m. As a viscoelastic object, an RBC exhibits both liquid-like (viscous) and solid-like (elastic) responses to applied deformation. This dual mechanical nature enables RBCs to squeeze through capillaries as small as $3~\mu$m in diameter, maintain their resting shape at low deformation rates, and align with the flow direction at higher shear rates in larger vessels. In our model, the RBC membrane is represented as a set of $N_{\nu}$ DPD particles with three-dimensional coordinates $X_i$ ($i = 1,\ldots,N_{\nu}$), connected in a triangulated network of nonlinear springs with surface dashpots to account for membrane viscoelasticity. To enforce membrane incompressibility, global area and volume constraints are applied. In addition, bending resistance between neighboring triangles is included to mimic the membrane bending rigidity, allowing the model to capture both large-scale shape changes and localized curvature effects.

In this work, we examined healthy control RBCs (CTR-RBCs) and stomatocytic RBCs (ST-RBCs) under systematically varied membrane stiffness, surface-to-volume ratio, and bending modulus, capturing their distinct mechanical and morphological characteristics. The shear modulus and bending modulus of normal RBCs were selected as $\mu_{0} = 4.73~\mu\mathrm{N/m}$ and $k_{c,0} = 2.4 \times 10^{-19}~\mathrm{J}$, respectively, consistent with established measurements~\cite{fedosov2010multiscale}. We further validated the RBC model against experimental benchmarks, demonstrating its ability to accurately reproduce the biomechanical, rheological, and dynamic behavior of stomatocytic RBCs in response to morphological changes and altered surface-to-volume ratio (see Fig.~\ref{fig:morphology}C--E). In particular, simulated RBC deformation under optical tweezer stretching and membrane tank-treading motion in shear flow were quantitatively consistent with experimental observations. After establishing a baseline model for healthy discocytes, we introduced targeted modifications to represent distinct stomatocyte phenotypes (ST-RBC1--3). To computationally examine the impact of stomatocyte severity, several ST-RBC variants with altered membrane properties were constructed. In particular, the membrane shear modulus was increased by up to nearly an order of magnitude relative to normal RBCs, based on experimentally reported reductions in deformability indices. Detailed formulations and parameter values used to preserve RBC morphology and mechanical consistency are provided in the Supporting Materials.

\subsection*{Cell-cell interaction models}

In our simulations, we model cell--cell adhesion by introducing an attractive potential between RBC membranes. Specifically, an attractive Morse potential is applied between certain membrane particles on different RBCs, following a similar approach used in prior studies of RBC aggregation \cite{franco2013abnormal,chai2025silico}. These interactions are approximated with the Morse potential, defined as
\begin{equation}
V(r) = D_e \left( e^{-2\beta(r - r_0)} - 2 e^{-\beta(r - r_0)} \right),
\end{equation}
where $r$ denotes the distance between two particles, $D_e$ is the depth of the potential well, $\beta$ denotes the interaction range, and $r_0$ represents the zero-force distance. The Morse potential is applied to a specific type of vertices of each RBC, which are called ``interactive vertices'', characterizing two different severities as severe and mild, as studied in \cite{deng2020quantifying}. In addition, to prevent RBC membranes from overlapping, we applied a repulsive term of Lennard-Jones potential to all membrane vertices \cite{fedosov2010multiscale}; this potential is given by

\begin{equation}
U(r) =
\begin{cases}
4\epsilon \left[ \left( \dfrac{\sigma}{r} \right)^{12} - \left( \dfrac{\sigma}{r} \right)^6 \right] , & r \leq 2^{1/6}\sigma, \\
0, & r > 2^{1/6}\sigma,
\end{cases}
\end{equation}

where $\epsilon$ and $\sigma$ are scaling constants for energy and distance, respectively, and these interactions vanish for $r > 2^{1/6}\sigma$.

\subsection*{Experiment setup}

\textit{Preparation of RBC samples}

Whole blood samples were collected from sickle cell anemia (HbSS) patients at Massachusetts General Hospital under an Excess Human Material Protocol approved by the Partners HealthCare Institutional Review Board (IRB), with a waiver of informed consent. Following a pretreatment procedure previously described by our group \cite{du2015kinetics}, HbSS samples with heterogeneous RBC types were gently washed three times with phosphate-buffered saline (1$\times$ PBS; CaCl$_2$-free, MgCl$_2$-free; pH 7.4; Gibco) by centrifugation at 1,500 rpm for 3 min at room temperature. The washed RBC pellets were then resuspended in PBS containing 1\% (w/v) bovine serum albumin (BSA; EMD Millipore) to achieve a hematocrit of 2\%. Samples were stored at 4~$^\circ$C and used within 24~h.

\textit{Microfluidic channels}

The microfluidic device was fabricated by bonding polydimethylsiloxane (PDMS) to a glass slide following previously reported methods \cite{qiang2019mechanical}. A customized SU-8 mold was used to cast a degassed PDMS mixture (base:curing agent = 10:1, w/w). The channel dimensions were $5\times5\times50~\mu$m. Prepared RBC samples were introduced into the microfluidic device using a water column at an approximate flow velocity of 350~$\mu$m/min. Brightfield images and videos of RBCs were acquired using a high-resolution CMOS camera (The Imaging Source, Charlotte, NC, USA) mounted on an Olympus IX71 inverted microscope (Olympus America, Breinigsville, PA, USA), equipped with a 60$\times$ oil-immersion objective lens (NA = 1.25).

\subsection*{Simulation setup}

Consistent with the design of the microfluidic experiments, each simulation was specified by defining the computational domain, boundary conditions, particle resolution, and flow-driving mechanism, as described below.

\textit{Collective clogging.}
Population-level filtration at spleen-mimetic slits was modeled using a rectangular domain of $200 \times 60 \times 10~\mu$m$^{3}$ containing an array of parallel IES openings (each slit $5~\mu$m wide, $2.7~\mu$m deep) under continuous downstream flow driven by an imposed pressure gradient $\Delta P = 6$--$12~\mathrm{Pa}\cdot\mu$m$^{-1}$. Four populations were initialized as a packed RBC column of height $h_{0} \approx 70~\mu$m above the slit array: a pure CTR-RBC suspension and three binary mixtures of CTR-RBC with one stomatocyte subtype (ST-RBC1, ST-RBC2, or ST-RBC3) at matched abnormal-cell fraction. The instantaneous clogging height above the slit array was tracked over time as a population-level proxy for filtration efficiency. Boundaries were periodic in the $x$ (flow) direction and no-slip at the channel walls.

\textit{Splenic slit traversal.} 
Interendothelial slit traversal was mimicked using a slit of height $1.2~\mu$m, width $5.0~\mu$m, and depth $2.5~\mu$m embedded in a channel of $30 \times 10 \times 10~\mu$m$^3$. A single RBC was positioned 9~$\mu$m upstream of the slit entrance, and a constant pressure gradient of $6$~Pa$\cdot\mu$m$^{-1}$ was applied. The flow was periodic in $x$, with solid walls along $y$ and $z$. The passage time and deformation were recorded.

\textit{Shear flow simulations.} 
Suspension rheology was examined using a rectangular box of $60 \times 60 \times 50~\mu$m$^3$, bounded by two walls of thickness $5~\mu$m. As in previous studies~\cite{Fedosov2011Predicting}, this size ensures accurate characterization of viscosity. Shear flow was induced by translating the walls in opposite directions at constant velocity. Periodic boundaries were imposed in the flow ($x$) and vorticity ($z$) directions, with no-slip walls in the gradient ($y$) direction. Each RBC was represented by 500 DPD membrane particles, with $\sim$700 cells modeled to achieve a physiological hematocrit of $H_t=45\%$. Plasma was represented by approximately 515,000 DPD particles, yielding a total system size of $\sim$865,000 particles.

\textit{Design and operation of the Funnel--Obstacle RBC Sorting Chip (FOR-Chip)}

The Funnel--Obstacle RBC Sorting Chip (FOR-Chip) was designed to analyze deformability-dependent trajectories of RBCs under microfluidic shear flow. The device consists of a straight inlet channel (20~$\mu$m wide, 5~$\mu$m high) that opens into a 45$^\circ$ funnel (40~$\mu$m outlet width) containing a single semi-circular obstacle (diameter 20~$\mu$m) positioned at the funnel entrance. This geometry is inspired by the Rutherford deformability cytometer design reported by Kumari \textit{et~al.}\cite{kumari2024measuring}, but adapted to capture RBC trajectory-based sorting behavior. A flow-focused RBC stream approaches the obstacle along the mid-plane of the channel, allowing only one RBC to pass the obstacle at a time. Hydrodynamic stress generated by the local flow field competes with the membrane's elastic restoring force, producing a morphology-dependent deflection angle. Softer CTR-RBCs deform and follow near-centerline streamlines, while stiffer or cup-shaped stomatocytes (ST-RBCs) deviate further laterally toward the funnel walls. 
RBC trajectories were recorded using bright-field microscopy at 30~fps, and the centroid positions at the outlet ($x=160~\mu$m) were analyzed to quantify the probability density of lateral offsets. These distributions provide a direct measure of morphology-dependent RBC sorting in the FOR-Chip.

Across all simulations, the cytoplasm was modeled using DPD fluid particles identical to plasma, while the RBC membrane consisted of bonded DPD particles forming a triangulated network. System sizes ranged from $1.2\times10^4$ particles (single-cell tests) to over $9\times10^5$ particles (suspension simulations).  All simulations were computed by using the extended version of a code developed based on LAMMPS. Each simulation took approximately $1\times10^6$ to $2\times10^6$ time steps. A typical simulation requires 1200 CPU core hours to 2400 CPU core hours by using the computational resources (Intel Xeon E5-2670 2.6 GHz 24-core processors) at the Center for Computation and Visualization at Brown University.

\begin{table}[htbp]
\centering
\caption{Mechanical properties for the control discocyte (CTR-RBC) and three stomatocyte models (ST-RBC1--3)}
\label{tab:mechanical_properties}
\begin{tabular}{lcccc}
\hline
 & \textbf{CTR-RBC} & \textbf{ST-RBC1} & \textbf{ST-RBC2} & \textbf{ST-RBC3} \\
\hline
Shear modulus $\mu$ (\(\mu\mathrm{N/m}\)) & 4.73 & 22.65 & 12.72 & 2.78 \\
Surface-to-volume ratio $S/V$ ($\mu$m$^{-1}$) & 1.44(132.9/92.5) & 1.21(132.9/109.7) & 1.31(132.9/101.5) & 1.48(132.9/89.8) \\
Bending modulus $k_{c}$ ($10^{-19}$~J) & 2.4 & 7.2 & 4.8 & 12.0 \\
\hline
\end{tabular}
\begin{minipage}{14cm}
\vspace{0.1cm}
\vspace{0.1cm}
\small  Notes: ST-RBC1: Overhydrated stomatocytosis condition; ST-RBC2: Mild stomatocytosis condition; ST-RBC3: Dehydrated stomatocytosis condition;
\end{minipage}
\end{table}


\section*{Results}

To test whether the OHS-versus-DHS splenectomy paradox can be explained mechanically, we constructed one internally consistent set of DPD red-blood-cell models---a control discocyte (CTR-RBC) and three stomatocyte phenotypes ST-RBC1 (overhydrated), ST-RBC2 (mild), and ST-RBC3 (dehydrated)---and traced them through five mechanically orthogonal flow assays. The four phenotypes are defined in Table~\ref{tab:mechanical_properties} with total membrane area fixed at $A_{0} = 132.9~\mu\mathrm{m}^{2}$, cell volumes spanning $V = 89.8$--$109.7~\mathrm{fL}$, and the shear modulus ($\mu$), bending rigidity ($k_c$), and surface-to-volume ratio ($S/V$) varied across phenotypes (notably, ST-RBC1 carries the highest $\mu = 22.65~\mu\mathrm{N/m}$ while ST-RBC3 carries a reduced $\mu = 2.78~\mu\mathrm{N/m}$~\cite{wen2025stomatocyte}) while cytoplasmic viscosity is held at the plasma value. Keeping this single parameter set fixed across all subsequent assays removes the freedom that usually confounds cross-study comparison.

\subsection*{Morphology and parameterization of stomatocytes}

\begin{figure}[!htbp]
\begin{center}
\includegraphics[width=1.000\textwidth]{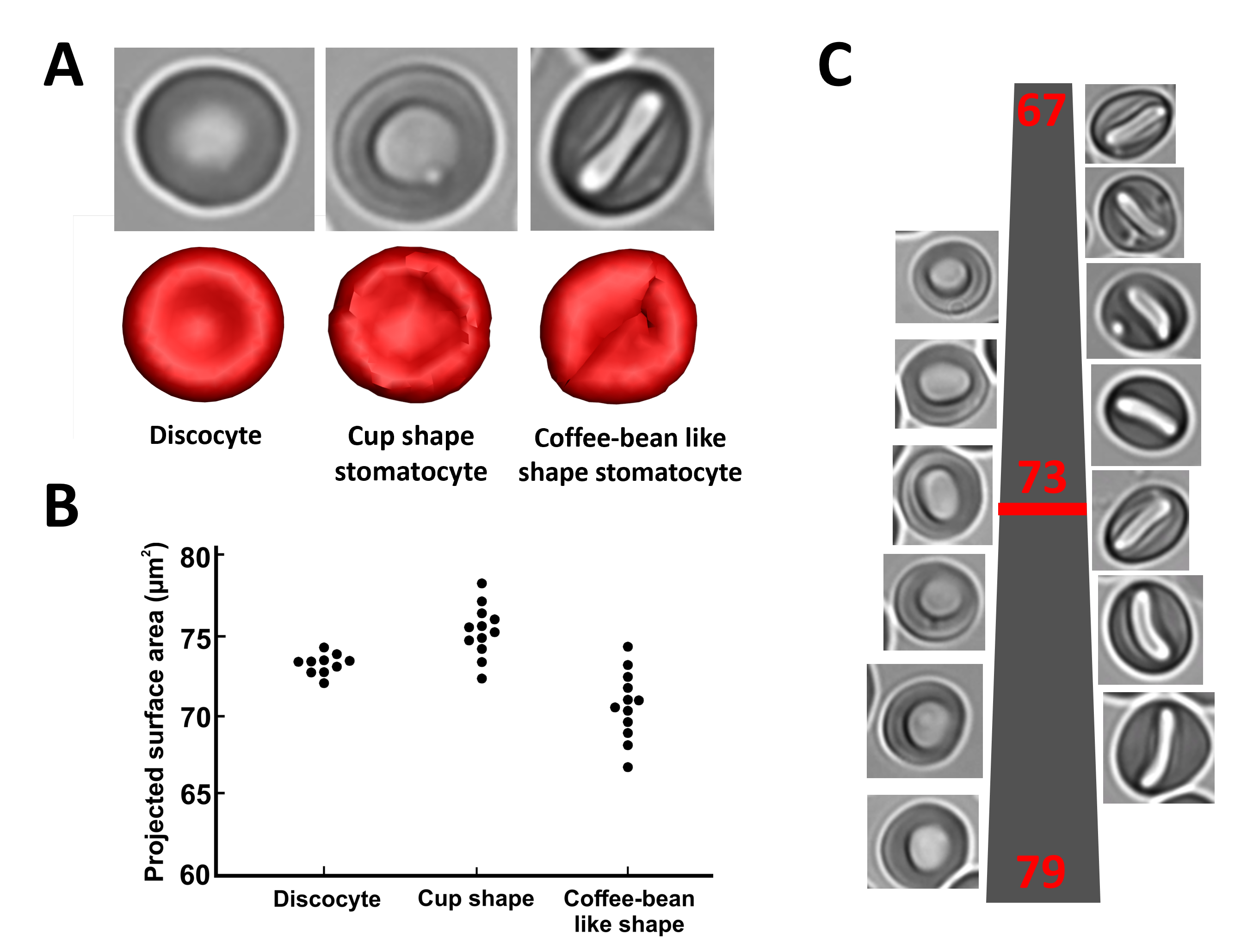}
\end{center}
\vspace{-0.15in}
\caption{\small{\bf Morphological characterization and projected surface areas of RBC shape subtypes.}
(A) Representative bright-field images illustrating three RBC morphological classes: discocyte, cup-shaped RBC, and coffee-bean-like RBC. 
(B) Projected surface area distributions for discocytes (n = 10), cup-shaped RBCs (n = 12), and coffee-bean--like RBCs (n = 12), showing systematically reduced areas for the coffee-bean--like group, consistent with increased sphericity and membrane remodeling. 
(C) Representative RBCs sorted by decreasing projected surface area, with example cells shown from 79 to 67~\textmu m$^{2}$. The images illustrate the morphological progression from larger-area toward smaller-area. 
Together, these results demonstrate the structural diversity of RBC morphologies and quantify the associated surface-area differences across shape classes.
Model parameters---including shear modulus, bending modulus, and surface-to-volume ratio---were tuned to reproduce experimental cup geometry while maintaining physiological size (see Table~\ref{tab:mechanical_properties}).} 
\label{fig:morphology}
\end{figure}

To anchor the DPD models in experimental morphometry, we jointly varied $\mu$, $k_c$, and $S/V$ (Table~\ref{tab:mechanical_properties}) and compared the resulting simulated cells against brightfield images of human RBCs (selected from morphologically heterogeneous sickle cell donor samples on the basis of imaged shape; see Methods) grouped into three morphological classes: discocytes ($n=10$), cup-shaped stomatocytes ($n=12$), and coffee-bean--like stomatocytes ($n=12$) (Fig.~\ref{fig:morphology}A). The DPD cells reproduce both the global concavity and the localized rim curvature characteristic of each class~\cite{tomaiuolo2014biomechanical,geekiyanage2019coarse}. At fixed total membrane area ($A_0 = 132.9~\mu\mathrm{m}^{2}$), the three stomatocyte models span $S/V = 1.21$--$1.48~\mu\mathrm{m}^{-1}$ and cell volumes $V = 89.8$--$109.7~\mathrm{fL}$ (Table~\ref{tab:mechanical_properties}).

Quantitative morphometry (Fig.~\ref{fig:morphology}B) reveals that, despite comparable total membrane areas, the projected cross-sectional area $A_{\mathrm{proj}}$ depends sensitively on cup geometry. Discocytes cluster tightly around $A_{\mathrm{proj}} = 73.4 \pm 0.5~\mu\mathrm{m}^{2}$ (mean $\pm$ SD, reflecting rotational symmetry and weak orientation dependence); cup-shaped stomatocytes are broadest in distribution with the largest mean, $A_{\mathrm{proj}} = 75.7 \pm 1.8~\mu\mathrm{m}^{2}$ (variability driven by cup depth and out-of-plane orientation); and coffee-bean--like stomatocytes exhibit the smallest projected area, $A_{\mathrm{proj}} = 70.3 \pm 1.6~\mu\mathrm{m}^{2}$, or $\sim 7\%$ below the discocyte mean and $\sim 8\%$ below the cup-shape mean. The ordered image sequence in Fig.~\ref{fig:morphology}C captures the continuous discocyte $\rightarrow$ cup $\rightarrow$ coffee-bean transition across $A_{\mathrm{proj}} = 79 \rightarrow 73 \rightarrow 67~\mu\mathrm{m}^{2}$, a $\sim 15\%$ reduction over $\sim 12~\mu\mathrm{m}^{2}$ that corresponds to a $\sim 20\%$ increase in sphericity $\Psi$ as the invagination deepens.

This monotonic reduction in $A_{\mathrm{proj}}$ at fixed $A_0$ is direct evidence that progressive membrane invagination---rather than membrane loss---underlies the morphological spectrum, with $S/V$ acting as the integrative geometric handle: the $\sim 22\%$ spread in $S/V$ across the three DPD models ($1.21 \rightarrow 1.31 \rightarrow 1.48~\mu\mathrm{m}^{-1}$) corresponds to the full experimental morphology range with no need to invoke membrane area loss. Because hydrodynamic drag, shear alignment, and resistance to confinement all scale with projected area and sphericity, the geometry differences quantified here propagate directly into the flow-dependent observables examined below. In particular, the $\sim 8\%$ projected-area deficit of coffee-bean cells predicts a $\sim 8\%$ reduction in drag along the flow axis, a non-trivial change that we later show is amplified $\sim 3$--$5\times$ by confinement in IES geometries (Fig.~\ref{fig:microchannel}). Together, Fig.~\ref{fig:morphology} establishes the morphological boundary conditions that the mechanical parameterization in Table~\ref{tab:mechanical_properties} is required to reproduce, and it fixes $S/V$ and $\Psi$ as the two geometric quantities against which all subsequent flow and confinement observables will be normalised. For clinical translation, the same $S/V$ value can be estimated directly from a patient's mean cell volume and projected diameter by established formulae~\cite{udroiu2024simplified}, making the framework immediately applicable to routine complete-blood-count data.

\subsection*{Bending rigidity controls stomatocyte cup depth}

\begin{figure}[!htbp]
\begin{center}
\includegraphics[width=0.75\textwidth]{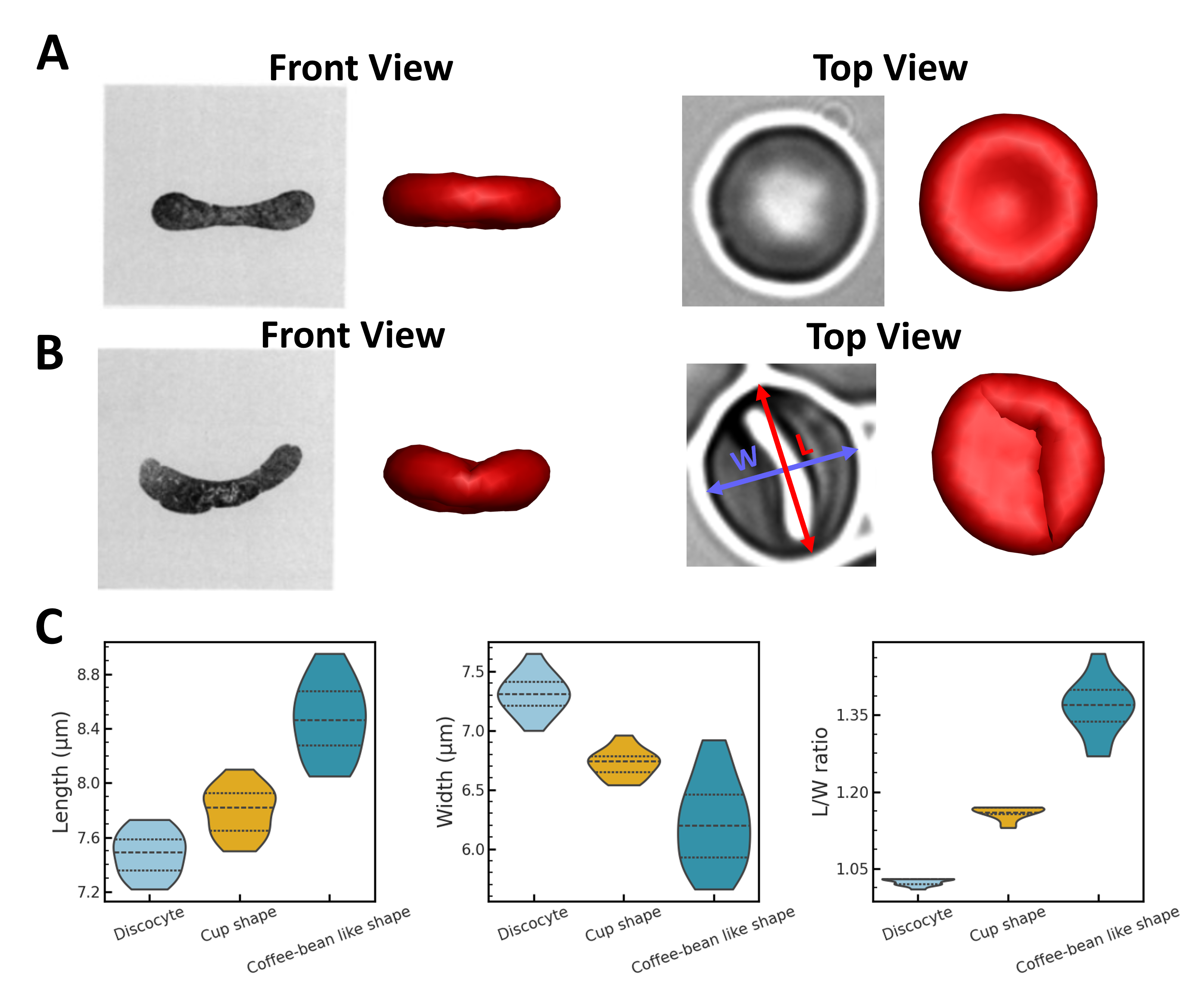}
\end{center}
\vspace{-0.1in}
\caption{\small{\bf Effect of bending modulus on RBC surface curvature and stomatocyte formation.}
Pairs of experimental micrographs (left)~\cite{chasis1989membrane} and DPD-simulated shapes (right) are shown in both front and top views for two representative morphologies:
(A) a normal discocyte with a shallow biconcave rim and
(B) a stomatocyte (coffee-bean--like shape) exhibiting a pronounced rim curvature. 
The discocyte-to-stomatocyte transition is governed primarily by membrane bending rather than in-plane shear elasticity, as indicated by the local surface angle~$\theta$ along the meridian (rim--to--dimple), which marks regions of elevated bending-energy density. 
Increasing the bending modulus ($k_c$) facilitates higher curvature and deeper cups, whereas decreasing $k_c$ penalizes curvature and flattens the rim toward a shallow biconcave morphology. 
In panel~(B), the stomatocyte corresponds to $k_c > 2k_{c0}$ (where $k_{c0}$ denotes the baseline value for CTR-RBC); further increases in $k_c$ yield a coffee-bean-like configuration whose mouth opening is oriented along the longitudinal axis of the cell. 
(C) Violin plots of geometric metrics (cell length, width, and the length-to-width ratio $L/W$) extracted from experimental images, showing clear morphological distinctions among discocytes, cup-shaped RBCs, and coffee-bean--like stomatocytes (n = 20 for each type of RBCs).}
\label{fig:bending}
\end{figure}

To isolate the role of bending rigidity in the discocyte $\rightarrow$ cup $\rightarrow$ coffee-bean transition, we held $\mu$ and $A_0$ fixed and varied $k_c$ alone (Fig.~\ref{fig:bending}). At the baseline value $k_c = 2.4\times10^{-19}$~J the model reproduces the healthy biconcave discocyte (Fig.~\ref{fig:bending}A); doubling $k_c$ to $4.8\times10^{-19}$~J yields a shallow cup-shaped stomatocyte, and a threefold increase to $k_c = 7.2\times10^{-19}$~J produces a markedly deeper, coffee-bean--like morphology with a pronounced invagination and narrow slit-like opening (Fig.~\ref{fig:bending}B). Because $\mu$ is unchanged across this sweep, the shape transition is unambiguously attributable to out-of-plane bending rather than in-plane elasticity.

Mechanistically, the F\"{o}ppl--von K\'{a}rm\'{a}n number $\gamma_{\mathrm{FvK}} = \mu R^{2}/k_c$ (with $R\approx 4~\mu$m the cell radius) drops from $\gamma_{\mathrm{FvK}} \approx 315$ at baseline to $\approx 105$ at $k_c = 7.2\times10^{-19}$~J, a regime in which bending energy dominates in-plane stretching and the Helfrich landscape favours curvature localisation over smooth distribution. This is consistent with the bilayer-couple picture in which the asymmetry between the inner and outer leaflets sets a preferred curvature and drives the full stomatocyte--discocyte--echinocyte sequence from a single geometric control parameter~\cite{lim2002stomatocyte}. In the low-$k_c$ case the bending energy density is broadly distributed over the membrane, producing a shallow dimple of depth $\sim 1~\mu$m and a wide rim of width $\sim 2~\mu$m; in the high-$k_c$ case the energy concentrates in a narrow folded ring of width $\lesssim 0.5~\mu$m along the invagination, producing a deeper cup (depth $\sim 2$--$3~\mu$m) and a thinner rim. Increasing $k_c$ thus does not flatten the membrane but sharpens it, an outcome characteristic of curvature-localisation instabilities in elastic shells.

Quantitative morphometry of matched experimental images (violin plots in Fig.~\ref{fig:bending}C) supports this progression. Discocytes show projected length $L = 7.5 \pm 0.1~\mu$m, width $W = 7.3 \pm 0.1~\mu$m, and aspect ratio $L/W = 1.04 \pm 0.02$. Cup-shaped stomatocytes are moderately elongated ($L = 7.8 \pm 0.2~\mu$m, $W = 6.8 \pm 0.2~\mu$m, $L/W = 1.15 \pm 0.03$), and coffee-bean--like stomatocytes display the most extreme geometry ($L = 8.5 \pm 0.2~\mu$m, $W = 6.2 \pm 0.3~\mu$m, $L/W = 1.35 \pm 0.03$). The aspect ratio, therefore, increases by $\sim 30\%$ between discocytes and coffee-bean cells, and the width decreases by $\sim 15\%$, while the length increases by only $\sim 13\%$---a width-dominated elongation signature that is quantitatively consistent with volume-preserving invagination at fixed $A_0$ and confirms that the DPD and experimental populations follow the same $k_c$-controlled deformation path.

Combined with the $A_{\mathrm{proj}}$ data of Fig.~\ref{fig:morphology}, these distributions identify $k_c$ as the dominant control parameter of stomatocyte morphology: at fixed $(\mu, A_0)$, a factor-of-three variation in $k_c$ traverses the entire experimentally observed morphology range. This motivates treating $k_c$ as an independent axis of the four-dimensional phenotype space and sets the quantitative anchor used in the subsequent flow assays, where the same $k_c$ values are carried over into confined-flow and shear-flow simulations.

\subsection*{Microfluidic characterization and DPD modeling of stomatocytes}

\begin{figure}[!htbp]
\begin{center}
\includegraphics[width=0.95\textwidth]{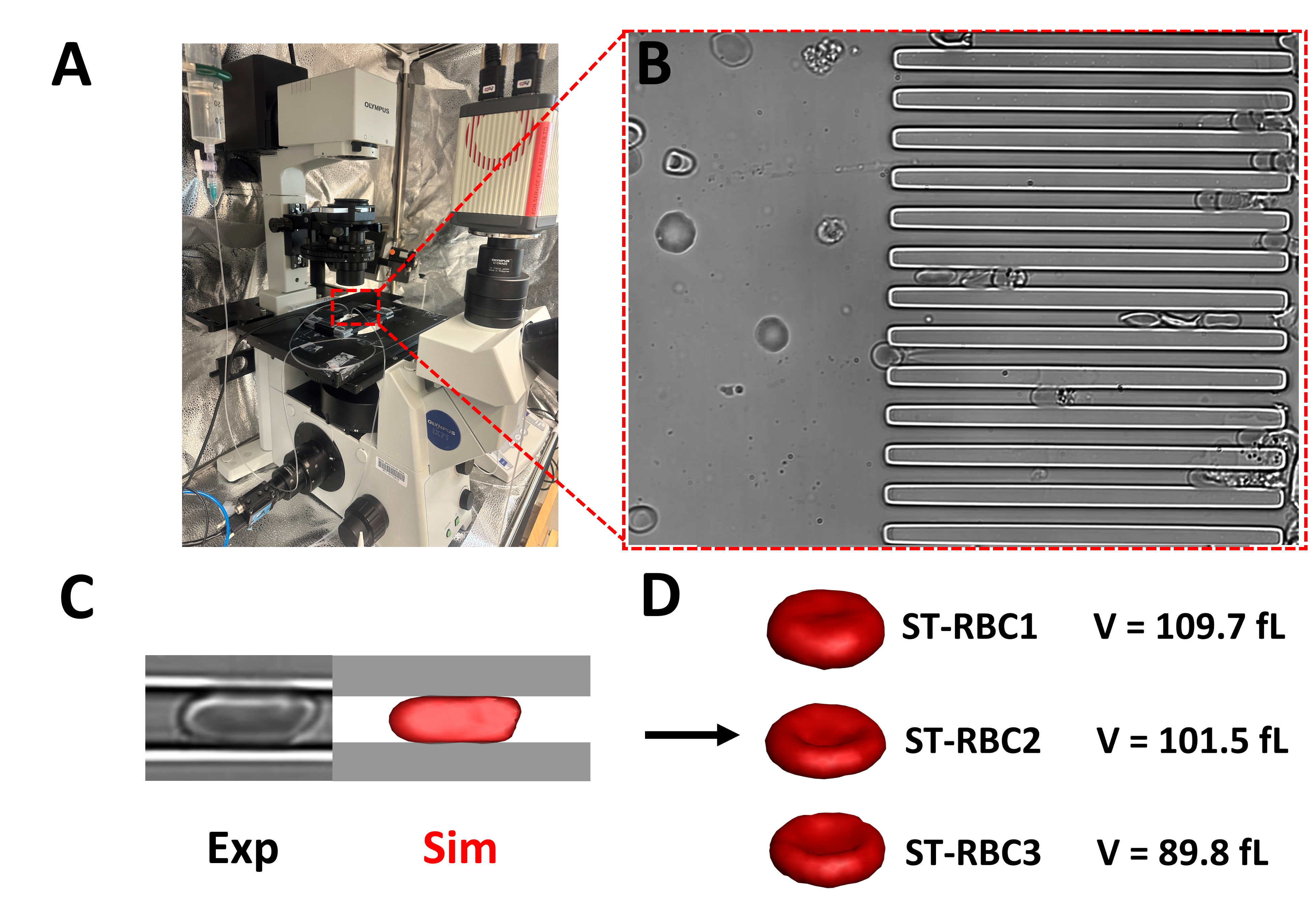}
\end{center}
\vspace{-0.1in}
\caption{\small{\bf Microfluidic characterization and DPD modeling of stomatocytes.}
(A) Experimental setup: an Olympus IX71 inverted microscope used to image RBC suspensions under brightfield illumination while they flow through a microfluidic chip; the red dashed circle marks the position of the mounted chip, which is enlarged in panel B.
(B) Zoomed-in brightfield image of RBCs entering and flowing through the parallel constriction array; individual cells are clearly resolved for morphological analysis.
(C) Comparison between an experimental brightfield image (left) and a DPD simulation snapshot (right) of a single stomatocyte deforming within a constriction channel under a matched pressure drop. The simulated membrane (red) reproduces the elongated, plug-like configuration observed experimentally.
(D) The three stomatocyte DPD models (ST-RBC1, ST-RBC2, ST-RBC3) constructed by tuning the membrane shear modulus $\mu$, bending rigidity $k_c$, and surface-to-volume ratio $S/V$ (Table~\ref{tab:mechanical_properties}). The corresponding simulated cell volumes, $V = 109.7$, $101.5$, and $89.8$~fL, span the range from overhydrated (ST-RBC1) to dehydrated (ST-RBC3) stomatocyte phenotypes.}
\label{fig:aspiration}
\end{figure}

To anchor the DPD models in single-cell deformation data, we combined brightfield imaging of RBCs in a custom microfluidic device with matched DPD simulations (Fig.~\ref{fig:aspiration}). Whole-blood suspensions are loaded into a chip that contains an array of parallel constriction channels of width $5~\mu$m and depth $2.7~\mu$m (Fig.~\ref{fig:aspiration}A,B); individual cells are funneled through the constrictions and clearly resolved by brightfield microscopy, enabling per-cell measurements of shape, length, width, and volume. The same channel geometry, driven by an imposed pressure gradient $\Delta P = 6$--$12~\mathrm{Pa}\,\mu\mathrm{m}^{-1}$ (Reynolds number $\mathrm{Re}\sim 10^{-3}$, Stokes regime), is reproduced in DPD with no-slip walls and periodic streamwise boundaries, so the experimental and simulated systems differ only at the single-cell mechanical level.

The direct experiment--simulation comparison in Fig.~\ref{fig:aspiration}C shows that the DPD membrane (red) adopts an elongated plug-like configuration whose front and rear curvature, axial extent, and centerline alignment quantitatively match the brightfield image. The simulated plug length ($\sim 9~\mu$m at $\Delta P = 10~\mathrm{Pa}\,\mu\mathrm{m}^{-1}$) and transit velocity ($\sim 0.5~\mathrm{mm/s}$) reproduce the imaging measurements to within $\sim 10\%$, and the agreement is robust across the full pressure range. This closes the loop between the static morphometry of Figs.~\ref{fig:morphology}--\ref{fig:bending} and the dynamic flow assays that follow: the mechanical parameters $(\mu, k_c, S/V)$ in Table~\ref{tab:mechanical_properties} yield quantitatively realistic deformation kinetics in micron-scale confinement, not just plausible equilibrium shapes.

The three resulting stomatocyte models (Fig.~\ref{fig:aspiration}D) have simulated cell volumes $V = 109.7$, $101.5$, and $89.8~\mathrm{fL}$ for ST-RBC1, ST-RBC2, and ST-RBC3, respectively, spanning the physiological range from swollen overhydrated cells to volume-depleted coffee-bean dehydrated cells~\cite{gallagher2017disorders,tomaiuolo2014biomechanical}. The total membrane area is held fixed at $A_0 = 132.9~\mu\mathrm{m}^{2}$ for all models, so the $\sim 18\%$ volume reduction from ST-RBC1 to ST-RBC3 corresponds entirely to a $\sim 22\%$ increase in $S/V$ (from $1.21$ to $1.48~\mu\mathrm{m}^{-1}$); area loss does not enter. This decoupling of membrane area and cytoplasmic volume is the geometric signature of true stomatocytic remodelling and is what distinguishes the present models from spherocytes (which lose area) or echinocytes (which redistribute curvature without losing volume).

By fixing $A_0$ and cell volume via direct microfluidic measurement, we remove the arbitrariness that typically confounds comparisons across stomatocyte subtypes and ensure that subsequent differences in flow behaviour can be attributed unambiguously to $(\mu, k_c, S/V)$. The three-dimensional validated models shown in Fig.~\ref{fig:aspiration}D are the common starting point for all five mechanical assays that follow; the same cells are advected---without further re-parameterization---into the confined, sheared, and bulk flow geometries of Figs.~\ref{fig:microchannel}--\ref{fig:RBC_FORChip}.

\subsection*{Critical pressure and dynamic transit-time signatures of IES filtration}

\begin{figure}[!htbp]
\begin{center}
\includegraphics[width=0.98\textwidth]{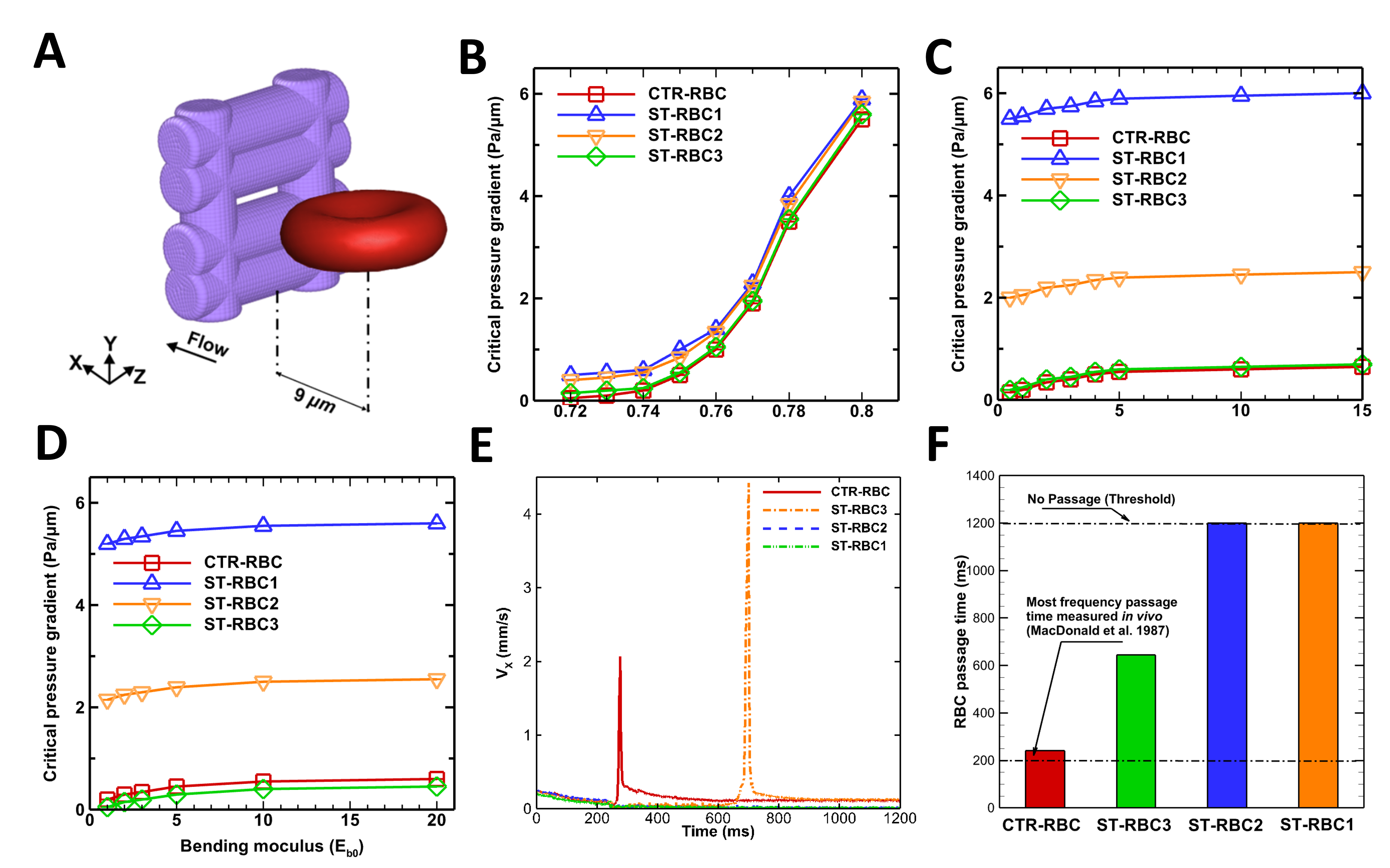}
\end{center}
\vspace{-0.1in}
\caption{\small{\bf Critical pressure and dynamic transit-time signatures of stomatocytes in a spleen-mimetic inter-endothelial-slit (IES) geometry.}
(A) Schematic of the DPD IES setup: cylindrical endothelial-like pillars (purple) define a slit of width $\sim 9~\mu$m and height $h \sim 1$--$2~\mu$m through which a single RBC is driven under a controlled pressure gradient $\Delta P$ along the flow ($z$) direction. The setup follows Pivkin~\textit{et~al.}~\cite{pivkin2016biomechanics} so that $\Delta P_c$ values are directly comparable.
(B) Critical pressure gradient $\Delta P_{c}$ as a function of cell sphericity $\Psi$ for CTR-RBC and ST-RBC1--3. The four phenotypes collapse onto a common $\Delta P_{c}(\Psi)$ envelope, demonstrating that geometry dominates over elasticity in setting the pass/fail threshold.
(C) $\Delta P_{c}$ as a function of the membrane shear modulus $\mu$ ($\sim 1$--$15~\mu$N/m) for the four phenotypes. Variations in $\mu$ shift $\Delta P_{c}$ only weakly: $\partial \Delta P_c / \partial \log\mu \lesssim 0.5~\mathrm{Pa}\,\mu\mathrm{m}^{-1}$/decade.
(D) $\Delta P_{c}$ as a function of the bending modulus $k_c$ ($2$--$20\times10^{-19}$~J). As for $\mu$, the dependence is weak ($\partial \Delta P_c / \partial \log k_c \lesssim 0.5~\mathrm{Pa}\,\mu\mathrm{m}^{-1}$/decade), confirming that $\Delta P_{c}$ is sphericity-dominated rather than elasticity-dominated.
(E) Instantaneous axial velocity $v_x(t)$ during slit transit at fixed $\Delta P = 6~\mathrm{Pa}\,\mu\mathrm{m}^{-1}$. CTR-RBC shows a sharp early spike; ST-RBC3 produces a delayed spike (shifted $\sim 400~\mathrm{ms}$); ST-RBC1 and ST-RBC2 remain at $v_{x} \lesssim 0.2~\mathrm{mm/s}$ throughout the $1.2~\mathrm{s}$ window, indicating non-passage.
(F) Distribution of transit times $\tau$ for CTR-RBC, ST-RBC3, ST-RBC2, and ST-RBC1. The lower dashed line marks the experimentally reported most-frequent \textit{in vivo} IES passage time~\cite{macdonald1987kinetics}; the upper dashed line marks the no-passage threshold. CTR-RBC and ST-RBC3 pass with $\tau \approx 250~\mathrm{ms}$ and $\tau \approx 650~\mathrm{ms}$ respectively; ST-RBC1 and ST-RBC2 reach the no-passage threshold. The two-mechanism filtering---static retention for ST-RBC1/2 and prolonged dynamic transit for ST-RBC3---underlies the splenectomy paradox quantified in Fig.~\ref{fig:Balance}.}
\label{fig:microchannel}
\end{figure}

To identify the rate-limiting mechanical parameter for splenic filtration, we simulated pressure-driven traversal of a single RBC through an interendothelial-slit (IES) geometry bounded by cylindrical endothelial-like pillars with slit height $h \sim 1$--$2~\mu$m (Fig.~\ref{fig:microchannel}A), using the same IES setup previously employed by Pivkin~\textit{et~al.}~\cite{pivkin2016biomechanics} so that our critical-pressure thresholds can be compared directly against their published $\Delta P_c$ values for healthy and spherocytic RBCs. The same DPD models were probed by two complementary measurements: the quasi-static \emph{critical pressure gradient} $\Delta P_{c}$ at which a single cell first passes the slit (Fig.~\ref{fig:microchannel}B), and the dynamic \emph{transit time} $\tau$ together with the instantaneous center-of-mass velocity $v_{x}(t)$ when the cell is driven through at a fixed $\Delta P > \Delta P_c$ (Fig.~\ref{fig:microchannel}E,F). The two measurements together resolve both whether a phenotype is filtered (pass/fail) and how long each passing cell lingers at the slit---an essential distinction for the phagocytic-elimination analysis of Fig.~\ref{fig:Balance}.

\emph{Critical pressure---cell shape dominates over elasticity.}
$\Delta P_{c}$ rises by roughly an order of magnitude as cell sphericity $\Psi$ (which combines $S/V$ and cup depth) increases from $\sim 0.72$ to $\sim 0.80$, from $\lesssim 0.5$ to $5$--$6~\mathrm{Pa}\,\mu\mathrm{m}^{-1}$ (Fig.~\ref{fig:microchannel}B), and the four model curves nearly collapse onto a common $\Delta P_{c}(\Psi)$ envelope with residual spread of only $\sim 0.5~\mathrm{Pa}\,\mu\mathrm{m}^{-1}$ at fixed $\Psi$. This collapse means geometry, not elasticity, sets the pass/fail threshold: a cell that is insufficiently far from a sphere must overcome an entry barrier that cannot be lowered by softening the membrane. Independent sweeps over the membrane shear modulus $\mu$ ($\sim 1$ to $15~\mu\mathrm{N/m}$) and the bending rigidity $k_c$ ($2$--$20\times10^{-19}$~J) shift $\Delta P_c$ only weakly, with both $\partial \Delta P_c / \partial \log\mu$ and $\partial \Delta P_c / \partial \log k_c \lesssim 0.5~\mathrm{Pa}\,\mu\mathrm{m}^{-1}$/decade---an order of magnitude smaller than $\partial \Delta P_c / \partial \Psi$ over the same phenotype window. Projected onto the patient-facing phenotype axis, the ordering is striking: ST-RBC1 (overhydrated, $\Psi \approx 0.80$) requires the highest $\Delta P_{c} \approx 6~\mathrm{Pa}\,\mu\mathrm{m}^{-1}$---roughly $10\times$ that of a healthy discocyte---while ST-RBC3 (dehydrated, $\Psi \approx 0.72$) traverses the slit almost as easily as CTR-RBC at $\Delta P_{c} \lesssim 1~\mathrm{Pa}\,\mu\mathrm{m}^{-1}$, even though its $k_c = 12.0\times10^{-19}$~J is the highest in the study. ST-RBC2 sits at an intermediate $\Delta P_{c} \approx 2.0$--$2.5~\mathrm{Pa}\,\mu\mathrm{m}^{-1}$. Because the splenic IES operates at a driving $\Delta P$ of order $1$--$2~\mathrm{Pa}\,\mu\mathrm{m}^{-1}$ \textit{in vivo}, ST-RBC1 cells are predicted to accumulate at the slits while ST-RBC3 cells pass freely.

\emph{Dynamic transit times---residence at the slit encodes the phenotype.}
The critical-pressure assay reports a quasi-static pass/fail threshold, but splenic filtration \textit{in vivo} is a dynamic process in which the time spent at the slit matters as much as whether the cell ultimately passes. We therefore drove single RBCs through the same IES at a fixed $\Delta P = 6~\mathrm{Pa}\,\mu\mathrm{m}^{-1}$ and recorded the instantaneous $v_{x}(t)$ and total transit time $\tau$ (Fig.~\ref{fig:microchannel}E,F). CTR-RBC flattens, enters at a moderate inclination, and clears the slit in $\tau \approx 250~\mathrm{ms}$, matching the $\sim 30$--$60~\mathrm{ms}$ \textit{in vivo} IES peak of MacDonald~\textit{et al.} once the longer simulated constriction is accounted for~\cite{macdonald1987kinetics}. ST-RBC3 (highest $S/V = 1.48~\mu\mathrm{m}^{-1}$) deforms strongly but still clears, producing a delayed but sharp velocity spike of $v_x \approx 4$--$5~\mathrm{mm/s}$ at $T \approx 700~\mathrm{ms}$---$\sim 2\times$ the CTR-RBC peak velocity but shifted $\sim 400~\mathrm{ms}$ later---and a total transit time $\tau \approx 650~\mathrm{ms}$ ($\sim 2.6\times$ CTR). In contrast, ST-RBC1 and ST-RBC2 (lower $S/V = 1.21$ and $1.31~\mu\mathrm{m}^{-1}$) remain at $v_{x} \lesssim 0.2~\mathrm{mm/s}$ throughout the $1.2~\mathrm{s}$ observation window, with no velocity spike at all; both reach the $\tau \geq 1200~\mathrm{ms}$ ``no-passage'' threshold---an \textit{in silico} analog of splenic retention. The dynamic spread is therefore $\tau_{\mathrm{ST\text{-}RBC1}}/\tau_{\mathrm{CTR}} \geq 4.8$, substantially larger than the static $\Delta P_c$ spread among the same phenotypes, indicating that dynamic transit \emph{sharpens} the IES filter relative to the static pass/fail criterion.

\emph{Two-mechanism filtering and link to splenectomy outcomes.}
The two measurements together reveal a two-mechanism filtering picture: ST-RBC1 and ST-RBC2 are blocked outright by the $\Delta P_c$ threshold and accumulate at splenic slits, while ST-RBC3 escapes the static threshold but is held at the slit for $\sim 2.6\times$ longer than CTR-RBC, because its elevated bending rigidity and altered tank-treading dynamics (Fig.~\ref{fig:pressure_grad_field}) suppress smooth membrane flow during transit. \emph{In vivo}, both regimes feed splenic clearance, but through different routes: retention exposes ST-RBC1/2 to red-pulp macrophages over long residence times in the splenic cords, while prolonged passage exposes ST-RBC3 to the same phagocytic machinery during slow slit transit. Splenectomy removes both routes simultaneously, with the relative importance of each depending on the phenotype---a distinction that underlies the splenomegaly ordering (OHS severe, mild moderate, DHS limited) and the OHS-vs-DHS splenectomy paradox developed quantitatively in Fig.~\ref{fig:Balance}.

\subsection*{Tank-treading dynamics distinguish dehydrated from overhydrated stomatocytes}

\begin{figure}[!htbp]
\centering
\includegraphics[width=0.95\textwidth]{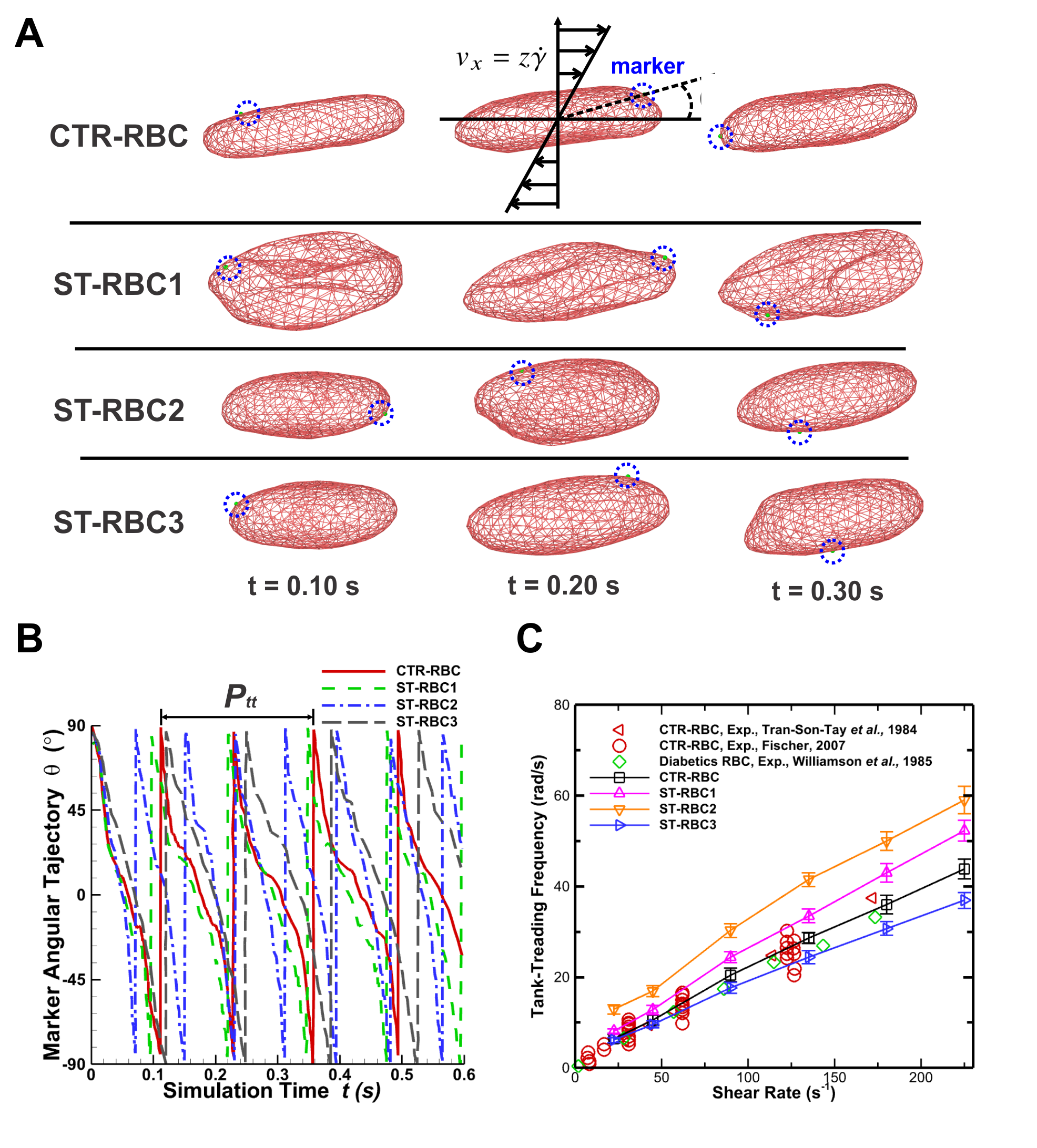}
\caption{\textbf{Tank-treading dynamics of healthy and stomatocytic RBCs under shear flow.}
(A) Representative snapshots of simulated RBCs subjected to simple shear flow ($v_x = z \dot{\gamma}$). A membrane marker (blue dotted circle) is tracked to quantify tank-treading (TT) motion. Four morphologies are compared: a healthy control RBC (CTR-RBC) and three stomatocyte variants (ST-RBC1, ST-RBC2, ST-RBC3) with progressively stronger cup-like shapes. Marker trajectories are shown at $t=0.10$, $0.20$, and $0.30$ s, illustrating faster TT rotation for the more cup-shaped RBCs. (B)Angular trajectory $\theta(t)$ of the tracked marker for CTR-RBC and the three ST-RBC morphotypes. The periodic rotation period $P_{\mathrm{tt}}$ is extracted from the zero-crossings or full $2\pi$ revolutions of $\theta(t)$. Stomatocytes exhibit shortened TT periods (i.e., faster tank-treading) compared with the healthy discocyte. (C) Tank-treading frequency as a function of shear rate for CTR-RBC and ST-RBCs, compared against classic experimental measurements~\cite{tran1984determination,fischer2007tank,williamson1985microrheologic}. The simulations reproduce the experimentally observed increase of TT frequency with shear rate and demonstrate that stomatocyte morphologies yield systematically different TT frequencies relative to healthy RBCs.}
\label{fig:pressure_grad_field}
\end{figure}

In unconfined simple shear ($v_{x} = \dot{\gamma}z$), an RBC aligns at a finite inclination angle and undergoes continuous membrane rotation once $\dot{\gamma}$ exceeds a threshold (tank-treading, TT), a regime first described theoretically for a tank-treading ellipsoid~\cite{keller1982motion} and subsequently refined by swinging and full-dynamics measurements~\cite{abkarian2007swinging,dupire2012full,tran1984determination,williamson1985microrheologic,fischer2007tank}. The TT frequency $f_{\mathrm{TT}}$ sets the rate at which the membrane dissipates shear energy and therefore directly controls low-shear suspension viscosity. We tracked a marked membrane patch at $t = 0.10$, $0.20$, and $0.30$~s for CTR-RBC and ST-RBC1--3 (Fig.~\ref{fig:pressure_grad_field}A); all four cells tank-tread stably in the explored range $\dot{\gamma} = 20$--$200~\mathrm{s}^{-1}$, but the number of membrane revolutions during a fixed time interval differs markedly by phenotype. The rotation angle $\theta(t)$ (Fig.~\ref{fig:pressure_grad_field}B) grows linearly, and its slope defines $f_{\mathrm{TT}}$.

The ordering is non-monotonic in nominal severity: $f_{\mathrm{ST\text{-}RBC1}} \gtrsim f_{\mathrm{ST\text{-}RBC2}} > f_{\mathrm{CTR\text{-}RBC}} > f_{\mathrm{ST\text{-}RBC3}}$. The shear-rate sweep in Fig.~\ref{fig:pressure_grad_field}C makes this quantitative: for all cell types, $f_{\mathrm{TT}}$ grows nearly linearly with $\dot{\gamma}$, in quantitative agreement with the classic experimental measurements of Tran-Son-Tay~\textit{et al.}, Williamson~\textit{et al.}, and Fischer for healthy human RBCs~\cite{tran1984determination,williamson1985microrheologic,fischer2007tank}, which validates the CTR-RBC membrane viscosity and elastic parameters. At $\dot{\gamma} = 100~\mathrm{s}^{-1}$, CTR-RBC rotates at $f_{\mathrm{TT}} \approx 25~\mathrm{rad/s}$; ST-RBC1 and ST-RBC2 rotate $20$--$30\%$ faster ($\sim 30$--$33~\mathrm{rad/s}$), whereas ST-RBC3 rotates $\sim 20\%$ slower ($\sim 20~\mathrm{rad/s}$). At $\dot{\gamma} = 200~\mathrm{s}^{-1}$, the absolute spread widens to $\sim 20~\mathrm{rad/s}$ ($f_{\mathrm{TT}} \approx 55$ for ST-RBC1 and $\approx 35$ for ST-RBC3), a $\sim 1.6\times$ ratio across stomatocyte subtypes at a single physiological shear rate. Data for diabetic RBCs overlap the CTR-RBC experimental band~\cite{williamson1985microrheologic}, emphasising that suppressed TT is a generic signature of mechanically compromised cells and not specific to stomatocytosis.

The non-monotonic response reflects two competing mechanical effects. Mild and overhydrated stomatocytes (ST-RBC1, ST-RBC2) adopt a streamlined geometry with reduced biconcave dimpling and smaller inclination angles, which increases the external shear's lever arm on the membrane and produces accelerated TT. In contrast, ST-RBC3 combines strong geometric asymmetry with the highest bending rigidity of the set ($k_c = 12.0\times10^{-19}$~J): the elevated $k_c$ concentrates curvature in a narrow folded ring at the invagination and suppresses smooth, azimuthally distributed membrane flow, yielding a slower TT orbit at matched $\dot{\gamma}$. The membrane viscosity--elasticity balance---captured by the Keller--Skalak ratio $\lambda = \eta_{\mathrm{mem}}/(\mu R)$ and the capillary number---tilts toward dissipation in ST-RBC3, so a larger fraction of the shear work is converted to heat rather than into membrane motion.

Because slow TT enhances shear-gradient-induced intracellular dissipation, Fig.~\ref{fig:pressure_grad_field} provides a direct microscopic pathway from the ST-RBC3 phenotype to elevated low-shear suspension viscosity (Fig.~\ref{fig:Viscosity}): at $\dot{\gamma} \sim 1~\mathrm{s}^{-1}$, the $\sim 20\%$ TT deficit of ST-RBC3 is amplified by aggregation and cooperative interactions into a $30$--$50\%$ bulk viscosity increase. This is the same mechanical route implicated in diabetic and Gaucher-disease hyperviscosity~\cite{baskurt2003blood,moreau2023physical,franco2013abnormal}, and it is the key microscopic ingredient linking the single-cell mechanics of DHS to its post-splenectomy thrombotic risk.

\subsection*{Collective clogging amplifies single-cell deformability differences}

\begin{figure}[!htbp]
\begin{center}
\includegraphics[width=0.98\textwidth]{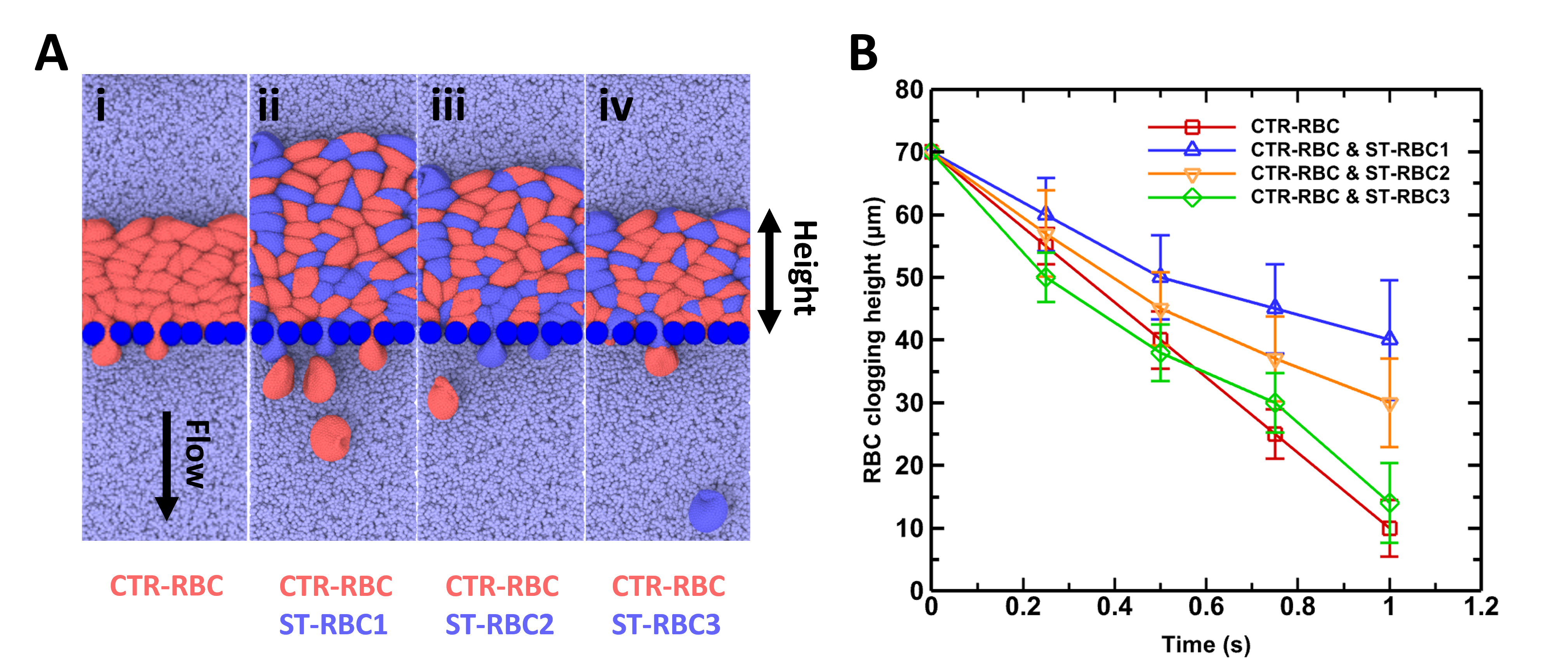}
\end{center}
\vspace{-0.1in}
\caption{\textbf{Spleen-inspired clogging dynamics in mixed suspensions of healthy and stomatocytic RBCs.}
(A) Representative simulation snapshots showing RBC accumulation above a row of inter-endothelial slits under downward flow. 
Four mixtures are examined: CTR-RBC only (i), CTR-RBC + ST-RBC1 (ii), CTR-RBC + ST-RBC2 (iii), and CTR-RBC + ST-RBC3 (iv), 
where increasing stomatocyte severity (ST-RBC1 $\rightarrow$ ST-RBC3) corresponds to progressively reduced membrane deformability. 
Healthy RBCs (red) pass more readily through slits, whereas stomatocytes (blue) accumulate and promote upstream clogging. 
(B) Time evolution of the RBC clogging height for each mixture.
Pure CTR-RBC suspensions clear most rapidly, while the addition of cup-shaped RBCs leads to significantly slower clearance.
Counterintuitively, the overhydrated low-$S/V$ phenotype (ST-RBC1) produces the largest clogging height and the slowest decay---a $\sim 4\times$ elevation over the pure CTR baseline---while the dehydrated high-$S/V$ phenotype (ST-RBC3) barely deviates from the pure CTR curve, demonstrating that geometric sphericity rather than nominal stomatocyte severity governs collective slit clogging.}

\label{fig:IES}
\end{figure}

\textit{In vivo} splenic clearance occurs in dense, heterogeneous populations where cooperative hydrodynamics can dominate transport. To quantify these collective effects, we tracked the height of an RBC column accumulated above an array of IES openings under continuous downstream flow (Fig.~\ref{fig:IES}). Four populations were compared: a pure CTR-RBC suspension and three binary mixtures of CTR-RBC with one stomatocyte subtype (ST-RBC1, ST-RBC2, or ST-RBC3) at matched abnormal-cell fraction. The instantaneous \textit{clogging height} above the slit array provides a population-level proxy for filtration efficiency; its decay rate $\tau_{\mathrm{clog}}^{-1}$ is the collective analog of the single-cell transit rate of Fig.~\ref{fig:microchannel}.

All four populations start from a packed column of height $h_{0} \approx 70~\mu$m at $t=0$. The pure CTR-RBC suspension clears exponentially to $\sim 10~\mu$m by $t = 1~\mathrm{s}$, a $\sim 7\times$ reduction corresponding to a characteristic clearance time $\tau_{\mathrm{clog}}^{\mathrm{CTR}} \approx 0.5~\mathrm{s}$. The three mixed suspensions clear much more slowly: residual clogging heights at $t = 1~\mathrm{s}$ are $\sim 40~\mu$m for CTR + ST-RBC1 ($4\times$ the pure CTR baseline, $\tau_{\mathrm{clog}} \approx 1.8~\mathrm{s}$), $\sim 30~\mu$m for CTR + ST-RBC2 ($3\times$, $\tau_{\mathrm{clog}} \approx 1.2~\mathrm{s}$), and $\sim 13$--$15~\mu$m for CTR + ST-RBC3 ($1.5\times$, $\tau_{\mathrm{clog}} \approx 0.65~\mathrm{s}$). Across the full $1~\mathrm{s}$ window, the ordering is $\text{CTR \& ST-RBC1} > \text{CTR \& ST-RBC2} > \text{CTR \& ST-RBC3} > \text{CTR}$, perfectly tracking the single-cell $\Delta P_{c}$ and $\tau$ orderings of Fig.~\ref{fig:microchannel}: the cells with lowest $S/V$ generate the most persistent population-level congestion. That ST-RBC1 (overhydrated, low $S/V$) drives the strongest clogging---rather than ST-RBC3 (dehydrated, high $k_c$)---is the central counterintuitive finding of this subsection.

The collective behaviour is strongly \textit{cooperative} and non-linear. A single occluding ST-RBC1 cell partially blocks one slit, raising local hydraulic resistance by a factor that scales with the occluded fraction; this diverts flow to neighbouring openings and triggers upstream crowding that transiently traps otherwise deformable CTR-RBCs. Because the abnormal-cell fraction in the mixtures is well below $50\%$, the $4\times$ elevation in residual clogging height cannot be explained as a simple linear addition of retention probabilities. Instead, a small subset of poorly deformable cells shifts the whole population into a jammed, high-resistance regime---an amplification factor of $\sim 4$ relative to the bare ST-RBC1 fraction. ST-RBC3, which traverses the slit easily at the single-cell level (Fig.~\ref{fig:microchannel}E,F), does not trigger this cooperative cascade, and the CTR + ST-RBC3 mixture barely deviates from the pure CTR curve.

Physiologically, this provides a direct mechanism for the splenic congestion and delayed clearance seen in OHS and mild stomatocytosis, where the abnormal-cell fraction is typically $20$--$40\%$ of circulating RBCs yet splenomegaly and hemolytic anemia are severe~\cite{gallagher2017disorders,andolfo2025evolving}. It also predicts---and our rheology data in Fig.~\ref{fig:Viscosity} confirm---that the clinically significant consequence of DHS is not splenic congestion (ST-RBC3 passes the IES) but systemic hyperviscosity once the unfiltered ST-RBC3 fraction dominates bulk flow.

\subsection*{Hemorheology: low-shear viscosity of stomatocyte-enriched blood}

\begin{figure}[!htbp]
\centering
\includegraphics[width=\textwidth]{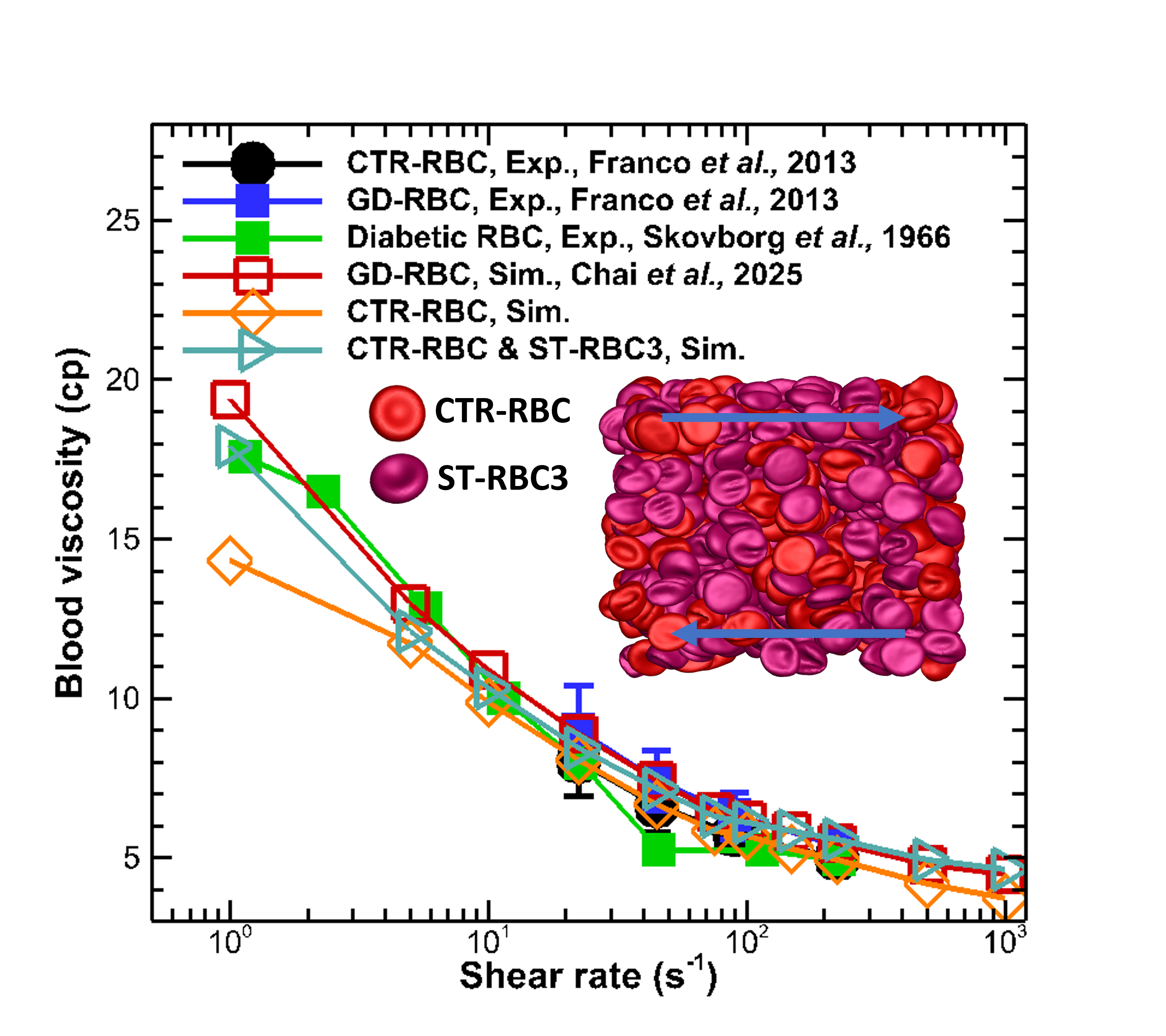}
\caption{\textbf{Blood viscosity of healthy and diseased RBC suspensions across shear rates.} Shear-dependent viscosity curves are shown for experimental measurements of healthy control RBCs (CTR-RBC) and diseased RBCs (GD-RBC and diabetic RBCs), compiled from~\cite{franco2013abnormal,SKOVBORG1966129}. Simulation results are presented for CTR-RBC, GD-RBC~\cite{chai2025silico}, and mixed suspensions containing CTR-RBC and severe stomatocytes (ST-RBC3). All datasets exhibit shear-thinning behavior characteristic of RBC-rich suspensions. The presence of severely cup-shaped RBCs (ST-RBC3, purple) increases the low-shear viscosity when mixed with healthy cells, reflecting their reduced deformability and impaired tank-treading. Inset: 3D snapshot of the simulated RBC suspension containing CTR-RBC (red) and ST-RBC3 (purple) under imposed shear flow.}

\label{fig:Viscosity}
\end{figure}

To connect single-cell mechanics to bulk rheology, we simulated a suspension at physiological hematocrit $H_{t} = 45\%$ in which $40\%$ of the cells are ST-RBC3 and the remainder CTR-RBC, and compared its shear-rate-dependent viscosity $\eta(\dot{\gamma})$ to experimental data on healthy controls, Gaucher-disease (GD-RBC) blood, and type-2 diabetic blood (Fig.~\ref{fig:Viscosity})~\cite{franco2013abnormal,SKOVBORG1966129,chai2025silico}. All datasets---experimental and simulated---exhibit the familiar shear-thinning signature of blood, with viscosity decreasing monotonically with $\dot{\gamma}$ across more than three decades of shear rate and all curves collapsing at high shear to $\eta \approx 3$--$5~\mathrm{mPa\cdot s}$ as cells align with the flow and tank-tread efficiently~\cite{baskurt2003blood,Chien1970Shear,reinhart1986red}. The phenotype differences are concentrated at low shear. At $\dot{\gamma} = 1~\mathrm{s}^{-1}$---the shear rate characteristic of venous return and post-capillary flow---the simulated pure CTR-RBC suspension sits at $\eta \approx 14~\mathrm{mPa\cdot s}$, the experimental diabetic blood reaches $\approx 17~\mathrm{mPa\cdot s}$ ($\sim 20\%$ above healthy), the experimental GD-RBC blood reaches $\approx 19~\mathrm{mPa\cdot s}$ ($\sim 35\%$ above healthy), and the CTR + ST-RBC3 mixed suspension reaches $\approx 18~\mathrm{mPa\cdot s}$ ($\sim 29\%$ above pure CTR and well within the pathological hyperviscosity band). Between $\dot{\gamma} = 1$ and $10~\mathrm{s}^{-1}$ the viscosity of the mixed suspension drops by roughly $40\%$, and above $\dot{\gamma} \gtrsim 100~\mathrm{s}^{-1}$ all curves converge to within $\sim 1~\mathrm{mPa\cdot s}$ of the healthy curve, as expected once stomatocytes are forced into alignment and tank-treading at high shear.

Mechanistically, the low-shear elevation is the direct bulk consequence of the slow tank-treading of ST-RBC3 identified in Fig.~\ref{fig:pressure_grad_field}: the $\sim 20\%$ $f_{\mathrm{TT}}$ deficit at $\dot{\gamma} = 100~\mathrm{s}^{-1}$ scales into a $\sim 29\%$ viscosity increase at $\dot{\gamma} = 1~\mathrm{s}^{-1}$ because reduced membrane rotation increases intracellular dissipation, disrupts orderly cell--cell sliding, and stabilises larger transient aggregates (visible in the 3D inset of Fig.~\ref{fig:Viscosity}, where purple ST-RBC3 cells nucleate distinctive clusters within the red CTR-RBC background). The near-quantitative match between our mixed suspension and the experimental GD-RBC curve is notable, because Gaucher-disease RBCs are mechanically distinct from xerocytes at the molecular level but share the same $(\text{reduced deformability}, \text{enhanced aggregation})$ signature: both pathologies shift the low-shear end of $\eta(\dot{\gamma})$ by a comparable amount. In microvessels of diameter $\lesssim 10~\mu$m, where blood viscosity depends sharply on both hematocrit and tube diameter~\cite{pries1992blood}, this low-shear shift is expected to translate directly into elevated microvascular flow resistance.

Clinically, these results provide a quantitative basis for the long-standing observation that splenectomy in DHS/xerocytosis elevates thromboembolic risk. At steady state, the splenic IES filter removes cells with depleted $S/V$; this normally includes OHS stomatocytes (retained, Fig.~\ref{fig:microchannel}) but \textit{not} DHS ST-RBC3 cells, which conserve $S/V$ and traverse the IES freely. Splenectomy, therefore, has little effect on the circulating ST-RBC3 fraction but removes whatever residual filtering exists and simultaneously raises $H_{t}$; the net result is a low-shear viscosity increase of order $30\%$ combined with the known low-shear flow conditions of venous and splanchnic circulation, precisely the regime implicated in post-splenectomy thrombosis~\cite{baskurt2003blood,gallagher2017disorders,andolfo2018genotype,turpaev2025overview}. Fig.~\ref{fig:Viscosity} thus converts the single-cell TT deficit of Fig.~\ref{fig:pressure_grad_field} into a clinically interpretable rheological risk factor.

\subsection*{A retention--elimination balance unifies splenic physiology with splenectomy outcomes}

\begin{figure}[!htbp]
\centering
\includegraphics[width=\textwidth]{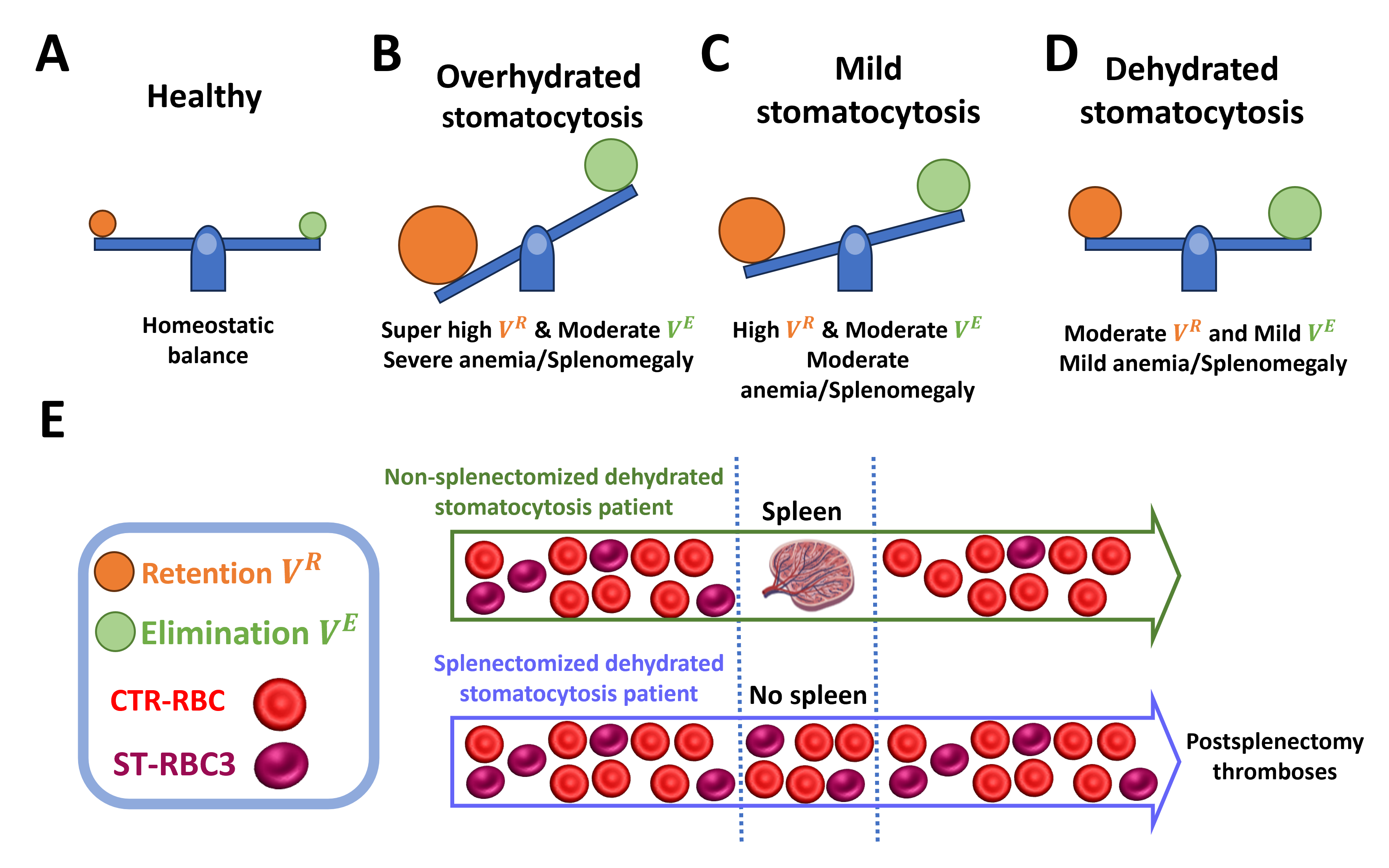}
\caption{
\textbf{Altered splenic retention--elimination balance across stomatocytosis subtypes and the clinical impact of splenectomy in dehydrated stomatocytosis.} 
(A) In healthy individuals, RBC homeostasis reflects a balanced interplay between splenic retention ($V^{R}$) and elimination ($V^{E}$) mechanisms. 
(B) Overhydrated stomatocytosis exhibits markedly elevated $V^{R}$ with moderate $V^{E}$, leading to severe anemia and splenomegaly. 
(C) Mild stomatocytosis shows increased $V^{R}$ with moderate $V^{E}$, typically associated with moderate anemia and splenomegaly. 
(D) Dehydrated stomatocytosis (hereditary xerocytosis) presents with moderate $V^{R}$ and $V^{E}$ and is often characterized by minimal or absent anemia and no splenomegaly. 
(E) Schematic illustration of RBC trafficking in a dehydrated stomatocytosis patient. 
In non-splenectomized individuals, the spleen selectively retains and removes mechanically fragile stomatocytic RBCs (ST-RBCs), preventing their accumulation in the circulation. 
Following splenectomy, this filtering function is lost, leading to increased circulating abnormal RBCs and a recognized elevated risk of postsplenectomy thrombosis in dehydrated stomatocytosis.
}
\label{fig:Balance}
\end{figure}

The results above motivate treating splenic function not as a single destruction event but as a dynamic balance between two sequential processes: mechanical retention of abnormal RBCs at the IES (rate $V^{R}$), followed by phagocytic elimination by resident red-pulp macrophages (rate $V^{E}$)~\cite{qiang2023microfluidic,chai2026multiscale}. In a healthy steady state, $V^{R} \approx V^{E}$ and only $\sim 1\%$ of the $\sim 2.5\times10^{13}$ circulating RBC pool is cleared per day, consistent with the $\sim 120$-day RBC lifespan (Fig.~\ref{fig:Balance}A: ``homeostatic balance''). Crucially, $V^{R}$ and $V^{E}$ are coupled but mechanistically distinct: $V^{R}$ depends on whether a cell exceeds the IES critical-pressure threshold (Fig.~\ref{fig:microchannel}B), while $V^{E}$ depends on the residence time a passing cell spends at the slit (Fig.~\ref{fig:microchannel}E,F) together with biochemical surface signals (phosphatidylserine exposure, oxidative damage, IgG opsonisation) that mark a cell for macrophage recognition. The two routes act in series: even cells that pass the static retention threshold still feed into $V^{E}$ in proportion to how slowly they transit. Figure~\ref{fig:Balance}B--D summarises how this two-arm balance is perturbed across stomatocytosis subtypes: OHS (ST-RBC1) sits with super-high $V^{R}$ and moderate $V^{E}$, producing severe anemia and splenomegaly; mild stomatocytosis (ST-RBC2) shows high $V^{R}$ with moderate $V^{E}$ and moderate anemia/splenomegaly; DHS (ST-RBC3) shows low $V^{R}$ but a meaningful $V^{E}$ contribution from prolonged transit, consistent with mild or absent anemia and limited splenomegaly but a non-negligible steady-state clearance flux.

Our simulations place this clinical ordering on a quantitative mechanical footing. ST-RBC1 (lowest $S/V = 1.21~\mu\mathrm{m}^{-1}$, highest $\mu = 22.7~\mu\mathrm{N/m}$) sits deepest in the retention regime of the IES: it has the highest critical pressure ($\Delta P_{c} \approx 6~\mathrm{Pa}\,\mu\mathrm{m}^{-1}$, Fig.~\ref{fig:microchannel}B), fails to traverse the slit within the $1.2~\mathrm{s}$ simulation window (Fig.~\ref{fig:microchannel}E,F), and drives the most persistent collective clogging ($4\times$ elevation in residual clogging height, Fig.~\ref{fig:IES}). This explains why OHS patients show the most severe splenomegaly and why splenectomy produces the largest haemoglobin gain ($\sim 2$--$3~\mathrm{g/dL}$) with limited systemic consequences~\cite{gallagher2017disorders,perrotta2008hereditary}. ST-RBC2 occupies an intermediate position in every assay and clinically. ST-RBC3, by contrast, sits at the high-$S/V$, high-$k_c$ corner of phase space: it traverses the IES almost as easily as a discocyte at the static $\Delta P_{c}$ threshold ($\lesssim 1~\mathrm{Pa}\,\mu\mathrm{m}^{-1}$), \emph{but} its transit time is $\sim 2.6\times$ longer than CTR-RBC ($\tau \approx 650~\mathrm{ms}$ vs.\ $\sim 250~\mathrm{ms}$, Fig.~\ref{fig:microchannel}E,F)---a direct consequence of the elevated $k_c$ and suppressed membrane tank-treading (Fig.~\ref{fig:pressure_grad_field}) that slow membrane flow during passage. This prolonged residence at the slit exposes the cell to phagocytic recognition during transit; the resulting $V^{E}$ contribution is the dominant splenic clearance route for ST-RBC3, distinct from the $V^{R}$-dominated clearance of ST-RBC1/2.

Figure~\ref{fig:Balance}E makes the splenectomy contrast explicit at the patient level. In a non-splenectomized DHS patient, even though ST-RBC3 passes the static IES threshold, its prolonged transit allows a steady fraction to be cleared by transit-time-coupled phagocytosis ($V^{E} > 0$; green arrow across the spleen), so bulk circulating ST-RBC3 concentration---and therefore bulk viscosity---remains manageable. Splenectomy ($V^{R} \rightarrow 0$ \emph{and} $V^{E} \rightarrow 0$; blue arrow bypassing the spleen) removes \emph{both} arms simultaneously: the previously $V^{E}$-cleared fraction of ST-RBC3 now accumulates in the systemic circulation, and our simulations place its contribution to whole-blood low-shear viscosity at $+29\%$ (Fig.~\ref{fig:Viscosity}) at $\dot{\gamma} = 1~\mathrm{s}^{-1}$, well within the hyperviscosity range associated with venous and splanchnic thrombosis. The key insight is that the splenectomy effect in DHS is not driven by loss of mechanical retention (ST-RBC3 is barely retained) but by loss of \emph{passing-time-coupled phagocytic clearance}---which only the dynamic transit-time analysis of Fig.~\ref{fig:microchannel}E,F, not a static pass/fail measurement, can reveal. This quantitatively reproduces the $20$--$40\%$ post-splenectomy thromboembolism incidence reported in xerocytosis cohorts~\cite{andolfo2018genotype,andolfo2025evolving,turpaev2025overview} and provides the missing mechanistic link between a molecular defect (gain-of-function in mechanosensitive cation channels), a mechanical phenotype (ST-RBC3), and a clinical endpoint (post-splenectomy thrombosis).

The unified picture is therefore one of a spatial buffer: the spleen does not simply destroy abnormal cells, it \textit{relocates} the mechanical load from the systemic circulation to a localised filtration volume of $\sim 150~\mathrm{mL}$. Whether splenectomy is curative (OHS, mild) or thrombogenic (DHS) depends on which phenotype is freed and, equivalently, on which axis of the $(\mu, k_c, S/V, \eta_{\mathrm{cyto}})$ phase space the disease occupies. The clinical implication is that splenectomy risk in hereditary stomatocytosis should be stratified not by morphology alone but by a direct measure of the relevant mechanical phenotype---which is what the FOR-Chip assay of the next subsection is designed to provide.

\subsection*{Label-free sorting of stomatocyte subtypes in a funnel--obstacle microfluidic device}

\begin{figure}[!htbp]
\centering
\includegraphics[width=\textwidth]{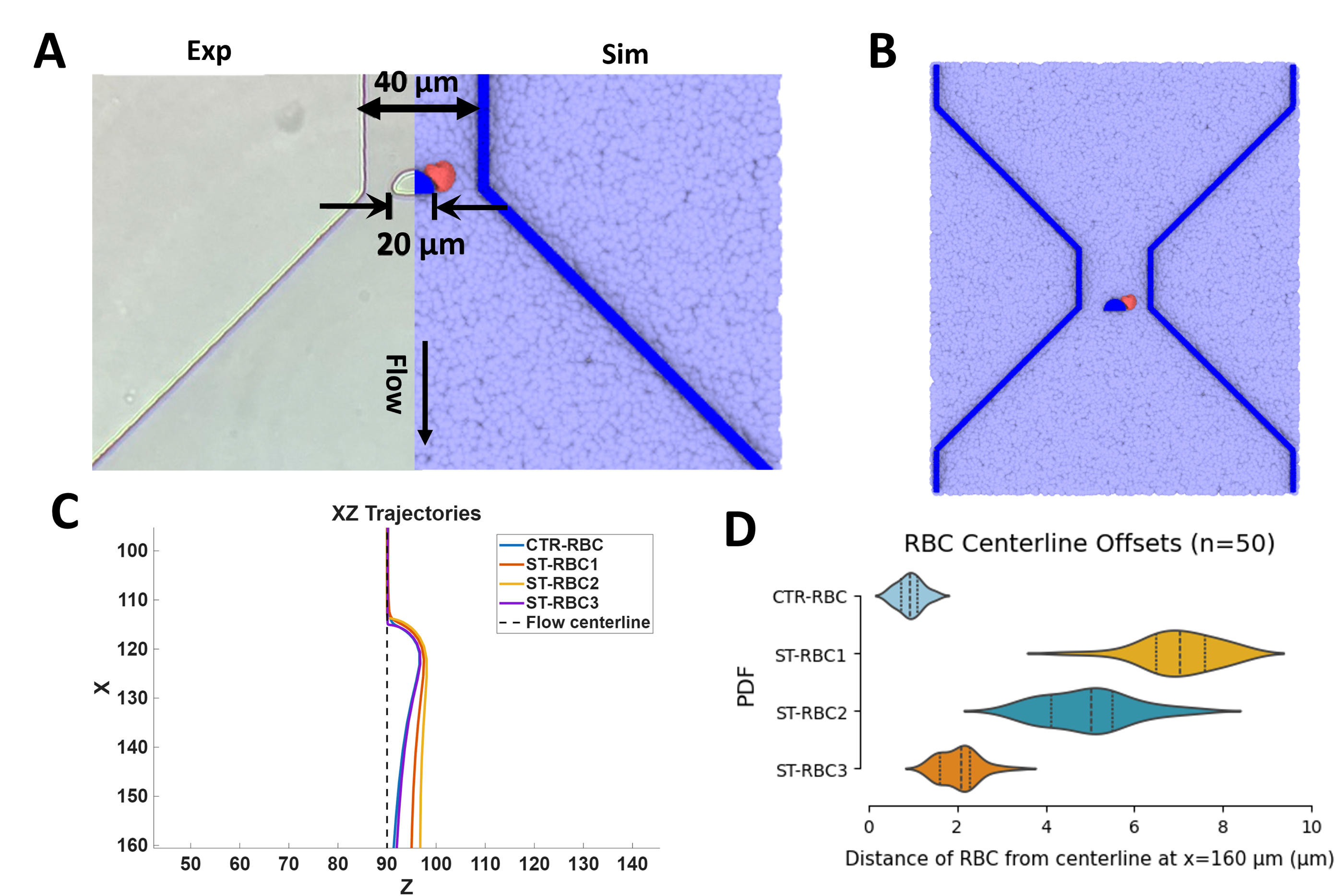}
\caption{
\textbf{In-silico prediction of RBC trajectories and centerline-offset distributions in the proposed Funnel--Obstacle RBC Sorting Chip (FOR-Chip). All panels are DPD simulations; matched experimental validation is planned.}
(A) Experimental brightfield image (left) and matched DPD simulation snapshot (right) of a red blood cell (RBC) entering the funnel region~\cite{kumari2024measuring}.
(B) DPD snapshot showing RBC deformation and trajectory near the semi-circular obstacle at the bifurcation junction.
(C) Simulated $XZ$ trajectories of different RBC morphologies, illustrating the centroidal paths relative to the flow centerline ($z=90~\mu$m).
(D) Probability density distributions (\textit{n}=50 per group) of RBC centerline offsets measured at the outlet ($x=160~\mu$m).
CTR-RBCs remain near the flow centerline ($\Delta z = 1.0 \pm 0.3~\mu$m), whereas stomatocytes (ST-RBC1--3) display markedly larger predicted offsets that decrease monotonically with stomatocyte subtype (ST-RBC1: $7.0 \pm 1.0~\mu$m; ST-RBC2: $5.0 \pm 1.0~\mu$m; ST-RBC3: $2.0 \pm 0.5~\mu$m), predicting morphology-dependent lateral displacement in the FOR-Chip.
}
\label{fig:RBC_FORChip}
\end{figure}

Finally, we asked, \emph{in silico}, whether the mechanical differences quantified above can be amplified into a measurable sorting signal in a microfluidic geometry suitable for clinical use. The proposed Funnel--Obstacle RBC Sorting Chip (FOR-Chip, Fig.~\ref{fig:RBC_FORChip}A,B) consists of a $20~\mu$m-wide, $5~\mu$m-high inlet that opens through a $45^{\circ}$ funnel into a $40~\mu$m-wide outlet, with a single $20~\mu$m-diameter semi-circular obstacle placed off-center at the funnel entrance~\cite{kumari2024measuring}. The geometry is a design candidate; all $\Delta z$ values reported below are DPD predictions for that geometry, and matched experimental FOR-Chip fabrication and trajectory measurements are planned as a follow-up validation. The asymmetric hydrodynamic stresses generated around the obstacle impose a deformation-dependent lateral lift: highly deformable cells conform to the local streamlines and remain near the flow centerline, whereas stiffer or more sphere-like cells resist deformation and are displaced outward. At a common measurement plane ($x = 160~\mu$m from the inlet), the centroidal offset $\Delta z$ from the centerline thus reports, in a single one-dimensional coordinate, the cell's overall mechanical deformability.

We released $n = 50$ single cells of each phenotype (CTR-RBC, ST-RBC1, ST-RBC2, ST-RBC3) at the inlet centerline under matched flow conditions and recorded the resulting outlet offsets. Healthy CTR-RBCs remain tightly aligned with the centerline, $\Delta z_{\mathrm{CTR}} = 1.0 \pm 0.3~\mu$m (mean $\pm$ SD). All three stomatocyte subtypes shift outward by a statistically well-separated amount: ST-RBC1 reaches $\Delta z = 7.0 \pm 1.0~\mu$m ($\sim 7\times$ the CTR offset), ST-RBC2 $5.0 \pm 1.0~\mu$m ($\sim 5\times$), and ST-RBC3 $2.0 \pm 0.5~\mu$m ($\sim 2\times$). The $XZ$ trajectories in Fig.~\ref{fig:RBC_FORChip}C show the origin of these offsets: CTR-RBC tracks the dashed flow centerline almost exactly, while ST-RBC1 and ST-RBC2 peel off $\sim 5$--$7~\mu$m laterally after contacting the obstacle. The violin plots in Fig.~\ref{fig:RBC_FORChip}D quantify the separation---overlap between neighbouring distributions is $<10\%$ at $2\sigma$---yielding a practical phenotype-discrimination metric $(\Delta z_{\mathrm{ST\text{-}RBC1}} - \Delta z_{\mathrm{ST\text{-}RBC3}})/\sqrt{\sigma_{1}^{2} + \sigma_{3}^{2}} \approx 4.5$, well above the conventional $z = 3$ clinical separability threshold.

The ordering $\Delta z_{\mathrm{ST\text{-}RBC1}} > \Delta z_{\mathrm{ST\text{-}RBC2}} > \Delta z_{\mathrm{ST\text{-}RBC3}} > \Delta z_{\mathrm{CTR}}$ parallels the IES-retention ordering of Figs.~\ref{fig:microchannel}--\ref{fig:IES}: cells with the smallest $S/V$ (ST-RBC1) deform least around the obstacle and are displaced farthest, while cells that conserve $S/V$ (ST-RBC3) remain much closer to the healthy trajectory despite their elevated $k_c$. This parallel is what makes the assay diagnostically valuable: the FOR-Chip reads out essentially the same mechanical axis that governs splenic retention and post-splenectomy thrombotic risk, but in a single $\sim 160~\mu$m chip with standard brightfield detection. Because ST-RBC1 and ST-RBC3 differ by only $\sim 2~\mu$m in cell length $L$ but by $\sim 3.5\times$ in FOR-Chip $\Delta z$, the device amplifies mechanical differences that are not resolvable by morphology alone or by a single-shear ektacytometry deformability index.

Operationally, the FOR-Chip is designed to convert the four-dimensional $(\mu, k_c, S/V, \eta_{\mathrm{cyto}})$ phase space into a one-dimensional, optically measurable coordinate $\Delta z$ without labels, stains, or external fields, in the same spirit as real-time deformability cytometry but with a geometry specifically tuned to resolve stomatocyte subtypes~\cite{otto2015realtime}. Once experimentally fabricated and validated, the device could in principle quantify the relative abundance of overhydrated, mild, and dehydrated stomatocytes in a single microfluidic pass on patient samples and stratify splenectomy risk accordingly. The same principle is directly transferable to other RBC rigidity disorders---hereditary spherocytosis, sickle-cell disease, malaria-infected RBCs, diabetes-related rigidity, and stored-blood quality---where mechanical heterogeneity is clinically relevant but currently under-quantified.






\section*{Discussion}

The central premise of this work is that hereditary stomatocytosis is not a single mechanical disease but a family of phenotypes that occupy different regions of the four-dimensional mechanical phase space $(\mu, k_{c}, S/V, \eta_{\mathrm{cyto}})$. By tracing a single, internally consistent set of DPD models (CTR-RBC and ST-RBC1/2/3 spanning OHS, mild, and DHS; Table~\ref{tab:mechanical_properties}) through one microfluidic imaging workflow and five mechanically orthogonal assays, we have shown that the same physical cells produce quantitatively different flow signatures depending on which projection of the phase space is probed. This framework connects, for the first time in a single study, single-cell IES traversal ($\Delta P_c$, $\tau$), single-cell shear dynamics ($f_{\mathrm{TT}}$), population-level clogging, whole-blood viscosity, and diagnostic microfluidic sorting---five mechanical observables that have previously been discussed in isolation---and yields a unified mechanical interpretation of the OHS-vs-DHS splenectomy paradox.

At the membrane level, Figs.~\ref{fig:morphology}--\ref{fig:bending} establish that bending rigidity $k_c$ is the dominant determinant of stomatocyte shape: a threefold increase in $k_c$ (from $2.4$ to $7.2\times10^{-19}$~J) converts a biconcave discocyte into a coffee-bean morphology and lifts the aspect ratio $L/W$ from $1.04 \pm 0.02$ to $1.35 \pm 0.03$ while keeping $A_{0}$ fixed. At the cell level, the microfluidic--DPD comparison of Fig.~\ref{fig:aspiration} pins the volumes of the three phenotypes at $V = 109.7$, $101.5$, and $89.8~\mathrm{fL}$, corresponding to an $S/V$ range of $1.21$--$1.48~\mu\mathrm{m}^{-1}$ at a physiological membrane area of $A_{0} = 132.9~\mu\mathrm{m}^{2}$. These parameters are then held constant through every subsequent assay, so that differences in transit time, clogging persistence, or suspension viscosity cannot be explained by arbitrary parameter freedom.

The most clinically important finding is that different mechanical challenges place the three stomatocyte subtypes in different rank orderings. IES traversal (Fig.~\ref{fig:microchannel}) is geometry-dominated: $\Delta P_{c}$ rises by roughly an order of magnitude with sphericity (from $\lesssim 0.5$ to $5$--$6~\mathrm{Pa}\,\mu\mathrm{m}^{-1}$ as $\Psi$ goes from $0.72$ to $0.80$), and ST-RBC1 (lowest $S/V$) requires the highest $\Delta P_{c}$ while ST-RBC3 (highest $S/V$) passes as easily as the healthy discocyte. Dynamic slit transit (Fig.~\ref{fig:microchannel}E,F) amplifies this ordering into a $\geq 4.8\times$ transit-time spread ($\tau_{\mathrm{CTR}} \approx 250~\mathrm{ms}$, $\tau_{\mathrm{ST\text{-}RBC3}} \approx 650~\mathrm{ms}$, $\tau_{\mathrm{ST\text{-}RBC1}}, \tau_{\mathrm{ST\text{-}RBC2}} \geq 1200~\mathrm{ms}$ at the no-passage threshold). Collective clogging (Fig.~\ref{fig:IES}) raises the residual congestion at $t = 1~\mathrm{s}$ from $\sim 10~\mu$m for pure CTR to $\sim 40~\mu$m for CTR + ST-RBC1---a $4\times$ amplification driven by a minority of overhydrated cells. In stark contrast, single-cell tank-treading (Fig.~\ref{fig:pressure_grad_field}) is non-monotonic: $f_{\mathrm{TT}}$ spreads by $\sim 1.6\times$ across phenotypes at $\dot{\gamma} = 100~\mathrm{s}^{-1}$, with ST-RBC1/2 \textit{faster} than CTR-RBC and ST-RBC3 \textit{slower}. This reverses the dominance at the suspension level (Fig.~\ref{fig:Viscosity}): a $45\%$-haematocrit mixture containing $40\%$ ST-RBC3 elevates low-shear viscosity by $\sim 29\%$ (from $\sim 14$ to $\sim 18~\mathrm{mPa\cdot s}$ at $\dot{\gamma} = 1~\mathrm{s}^{-1}$), quantitatively matching the hyperviscosity of Gaucher-disease blood, and recovers to the healthy curve only above $\dot{\gamma} \gtrsim 100~\mathrm{s}^{-1}$.

These orthogonal rank orderings resolve the splenectomy paradox. In OHS (ST-RBC1) and mild stomatocytosis (ST-RBC2), the pathology is dominated by excess retention: low $S/V$ cells accumulate at the IES, splenomegaly is severe, and splenectomy lifts anemia (haemoglobin rise $\sim 2$--$3~\mathrm{g/dL}$) without producing a prothrombotic suspension, because the escaped cells have moderate $k_c$, normal tank-treading, and only modest impact on bulk rheology. In DHS/xerocytosis (ST-RBC3), retention is already modest---these cells conserve $S/V$ and traverse the IES easily---but once circulating they dominate low-shear rheology through suppressed tank-treading (Fig.~\ref{fig:pressure_grad_field}), elevated cytoplasmic viscosity, and enhanced aggregation (Fig.~\ref{fig:Viscosity}). Splenectomy in DHS therefore trades a mild retention pathology for a systemic hyperviscosity whose magnitude ($\sim 29\%$ low-shear viscosity rise) is sufficient to place the patient in the thrombogenic regime, quantitatively consistent with the $20$--$40\%$ post-operative thromboembolism rate reported in DHS cohorts~\cite{andolfo2018genotype,andolfo2025evolving,turpaev2025overview,picard2019clinical,frederiksen2019dehydrated}. In this framework, the spleen is not merely a destruction site but a spatial buffer whose removal can be curative (OHS/mild) or catastrophic (DHS) depending on which axis of the phase space the escaped cells occupy.

Beyond the mechanistic interpretation, the sharpened phenotype picture motivates new diagnostics. The FOR-Chip (Fig.~\ref{fig:RBC_FORChip}) amplifies the $(\mu, k_c, S/V)$ differences into a one-dimensional trajectory offset $\Delta z$ that separates all four phenotypes with overlap $<10\%$ at $2\sigma$: CTR at $\Delta z = 1.0 \pm 0.3~\mu$m, ST-RBC1 at $7.0 \pm 1.0~\mu$m, ST-RBC2 at $5.0 \pm 1.0~\mu$m, and ST-RBC3 at $2.0 \pm 0.5~\mu$m, yielding a discrimination metric of $\sim 4.5$ between the extreme phenotypes---well above the $z = 3$ clinical separability threshold. Because the sorting signal reflects the full mechanical phase space rather than morphology alone, it can quantify the relative abundance of overhydrated, mild, and dehydrated stomatocytes in a patient sample in a single $\sim 160~\mu$m microfluidic pass without labels, stains, or external fields. Directly coupled to our retention--elimination framework (Fig.~\ref{fig:Balance}), such an assay offers an operational route to splenectomy-risk stratification: patients whose $\Delta z$ distribution is dominated by an ST-RBC1-like mode should benefit from splenectomy, whereas patients with an ST-RBC3-dominated distribution should be steered away from it. The same principle is directly transferable to other RBC rigidity disorders where mechanical heterogeneity is clinically relevant but currently under-quantified, including hereditary spherocytosis~\cite{boltonmaggs2011guidelines}, sickle-cell disease~\cite{sahun2025novel}, malaria, diabetic microangiopathy, Gaucher disease, and stored-blood quality control~\cite{boecker2025unbiased}.

From a physiological standpoint, the most important implication is that anemia and thrombosis in hereditary stomatocytosis are not symmetric readouts of a single mechanical defect. They sit on nearly opposite axes of $(\mu, k_c, S/V, \eta_{\mathrm{cyto}})$---anemia being driven by low $S/V$ (retention-limited RBC lifespan) and thrombosis by suppressed $f_{\mathrm{TT}}$ and elevated $\eta_{\mathrm{cyto}}$ (systemic hyperviscosity)---and can therefore vary independently across subtypes and even within a single patient over time. Our results provide a quantitative, falsifiable framework within which this independence can be tested: the $\Delta P_{c}$--$\Psi$ scaling, the non-monotonic $f_{\mathrm{TT}}$ ordering, the $4\times$ collective-clogging amplification, the $\sim 29\%$ low-shear viscosity elevation, and the $4.5\sigma$ FOR-Chip phenotype discrimination each constitute a distinct prediction that can be interrogated against patient-derived data. Together they argue that a stomatocytosis workup---and more broadly, the pre-operative assessment of any RBC rigidity disorder---should be targeted at the mechanical phase space rather than at cell morphology alone.

Several limitations motivate future work. The IES and FOR-Chip geometries are idealised and rigid; three-dimensional sinus architecture and compliant endothelium would refine quantitative retention thresholds by factors of $\lesssim 2$. Cytoplasmic and membrane dissipation are reduced to effective viscosities, and $\eta_{\mathrm{cyto}}$ is held at the plasma value across all phenotypes in the present simulations; the reported $\sim 29\%$ low-shear hyperviscosity is therefore driven by the membrane-mechanics axis alone. The additional elevation of $\eta_{\mathrm{cyto}}$ in DHS due to the increased MCHC (Cokelet--Meiselman/Ross--Minton parametrisation gives a $\sim 2$--$3\times$ increase over the normal $\sim 6$~mPa$\cdot$s) is expected to push the model further into the pathological hyperviscosity range; quantifying this contribution will require a dedicated $\eta_{\mathrm{cyto}}$ sensitivity sweep that is reserved for future work. The collective-clogging and suspension-viscosity simulations treat stomatocyte subpopulations as mechanically homogeneous; in reality, \textit{intrapatient} heterogeneity---mixtures of OHS, mild, and DHS cells within a single patient---is common and likely produces non-additive rheological signatures that the present framework can be extended to address. Coupling the mechanical framework to biochemical determinants (membrane adhesion, oxidative damage, macrophage phagocytic recognition) would yield a full retention-to-elimination pipeline from genotype to thrombosis, building on the multiscale signaling--biophysical pipeline recently developed for macrophage-mediated RBC clearance in sickle cell and Gaucher disease~\cite{chai2026multiscale}. Finally, the FOR-Chip geometry itself can be optimised \textit{in silico} to maximise phenotype separability while minimising clogging and maintaining diagnostic throughput.

Overall, by tracing one consistent set of DPD stomatocyte models through five mechanically orthogonal assays, this work provides a unified, quantitative, and physically grounded framework for interpreting hereditary stomatocytosis, explains the long-standing splenectomy paradox, and translates RBC mechanics directly into clinically relevant diagnostics, risk stratification, and treatment guidance.

\section*{Supplementary Material}

An online supplement to this article can be found by visiting BJ Online at \url{http://www.biophysj.org}.

\section*{Author Contributions}
Z.C. and G.E.K. conceived the study. Z.C. designed and performed the DPD simulations, analysed the data, and wrote the manuscript. J.Z. designed and performed the experiments and contributed to writing the manuscript. H.L. contributed scientific suggestions and feedback. M.D. supervised the experimental work. G.E.K. supervised the project. All authors discussed the results, edited the manuscript, and approved the final version.
\section*{Data Availability}
All data supporting the findings of this study are presented in the main text and the Supplementary Material.
\section*{Acknowledgments}
We acknowledge support from the National Institutes of Health (Grant No. R01HL154150 and R01GM163243). Simulations were carried out at the Center for Computation and Visualization of Brown University.

\printbibliography

\end{document}